\documentclass[11pt]{article}

\usepackage{JASA_manu}
\usepackage{hyperref}
\usepackage[utf8]{inputenc}
\usepackage{amsmath,amsfonts,amsthm}
\usepackage{graphicx}
\usepackage{xcolor}
\usepackage{natbib}
\usepackage[algo2e]{algorithm2e}
\usepackage{enumitem}
\usepackage{algorithm}

\begin{document}

\title{Outcome-Guided Disease Subtyping for High-Dimensional Omics Data}

\author{Peng Liu\\
Department of Biostatistics \\ University of Pittsburgh, Pittsburgh, PA 15261  \\ 
email: \texttt{pel67@pitt.edu} 
\and
Yusi Fang  \\
Department of Biostatistics \\ University of Pittsburgh, Pittsburgh, PA 15261  \\ 
email: \texttt{yuf31@pitt.edu} 
\and
Zhao Ren  \\
Department of Statistics \\ University of Pittsburgh, Pittsburgh, PA 15261  \\ 
email: \texttt{zren@pitt.edu} 
\and
Lu Tang \\
Department of Biostatistics \\ 
University of Pittsburgh, Pittsburgh, PA 15261\\
email: \texttt{lutang@pitt.edu} 
\and
George C. Tseng \\
Department of Biostatistics \\ 
University of Pittsburgh, Pittsburgh, PA 15261\\
email: \texttt{ctseng@pitt.edu} 
}

\maketitle

\newpage

\mbox{}

\begin{center}
\textbf{Abstract}
\end{center}
High-throughput microarray and sequencing technology have been used to identify disease subtypes that could not be observed otherwise by using clinical variables alone. The classical unsupervised clustering strategy concerns primarily the identification of subpopulations that have similar patterns in gene features. However, as the features corresponding to irrelevant confounders (e.g. gender or age) may dominate the clustering process, the resulting clusters may or may not capture clinically meaningful disease subtypes. This gives rise to a fundamental problem: can we find a subtyping procedure guided by a pre-specified disease outcome? Existing methods, such as supervised clustering, apply a two-stage approach and depend on an arbitrary number of selected features associated with outcome. In this paper, we propose a unified latent generative model to perform outcome-guided disease subtyping constructed from omics data, which improves the resulting subtypes concerning the disease of interest. Feature selection is embedded in a regularization regression. A modified EM algorithm is applied for numerical computation and parameter estimation. The proposed method performs feature selection, latent subtype characterization and outcome prediction simultaneously. To account for possible outliers or violation of mixture Gaussian assumption, we incorporate robust estimation using adaptive Huber or median-truncated loss function. Extensive simulations and an application to complex lung diseases with transcriptomic and clinical data demonstrate the ability of the proposed method to identify clinically relevant disease subtypes and signature genes suitable to explore toward precision medicine.

\vspace*{.3in}


\newpage

\section{Introduction}
\label{s:intro}
Many complex diseases were once considered a single disorder, within which all patients receive a uniform screening, diagnosis and treatment strategy. With better understanding of the underlying disease mechanisms, evidences have emerged to define novel subtypes of many complex diseases using clinical variables, selected biomarkers, imaging measurements, molecular profiling or genetic alterations, where the therapeutic plan can be tailored to each subtype to improve disease prognosis. In breast cancer, for example, four intrinsic subtypes (Lumina A, Lumina B, HER2-enriched and Basal-like) and a Normal Breast-like group were first identified in \cite{perou2000molecular} by cluster analysis of 42 patients based on microarray expression profile of 8102 genes and the result has been validated in many follow-up studies. Of the subtypes, Lumina A and Lumina B patients tend to have longer survival and lower recurrence rate, which require less aggressive treatment to reduce side effects. Basal-like (triple negative) tumors are often more malignant and have a poorer prognosis but can be successfully treated with certain combination of surgery, radiotherapy and chemotherapy. HER2-enriched patients can be treated with HER2-targeted therapy such as trastuzumab, which is surprisingly harmful to those in the Lumina subtypes. Subsequent tailored screening/prevention programs and novel treatment strategies from successful disease subtyping have decreased breast cancer mortality over the years \citep{jemal2009cancer}. Cluster analysis in high-dimensional omics data to characterize novel disease subtypes is an essential first step towards  precision medicine and is the focus of this paper. 

Classical clustering methods, such as hierarchical clustering, $K$-means clustering and Gaussian mixture model, have been widely used in the literature for disease subtyping. These methods are effective when the dimension of features is low and the clusters are well separated. The clustering task, however, becomes more challenging in high-dimensional omics data (e.g. thousands of genes in transcriptomic data) and the classical methods often fail to identify clinically meaningful clusters since they naively treat all features as equally important. Similar to most small-n-large-p problems, it is generally believed that only a small portion of features are relevant in the cluster characterization. A large amount of work has been devoted to dimension reduction and feature selection in cluster analysis, such as sparse principal component analysis or sparse factor analysis coupled with standard clustering \citep{zou2006sparse,bair2006prediction}, model-based clustering with variable selection \citep{tadesse2005bayesian,pan2007penalized} and sparse $K$-means \citep{witten2010framework}. Interested readers may refer to \cite{bouveyron2014model} for further reference.

\begin{figure}[!b]
	\resizebox{1\textwidth}{!}{\includegraphics[scale=0.7]{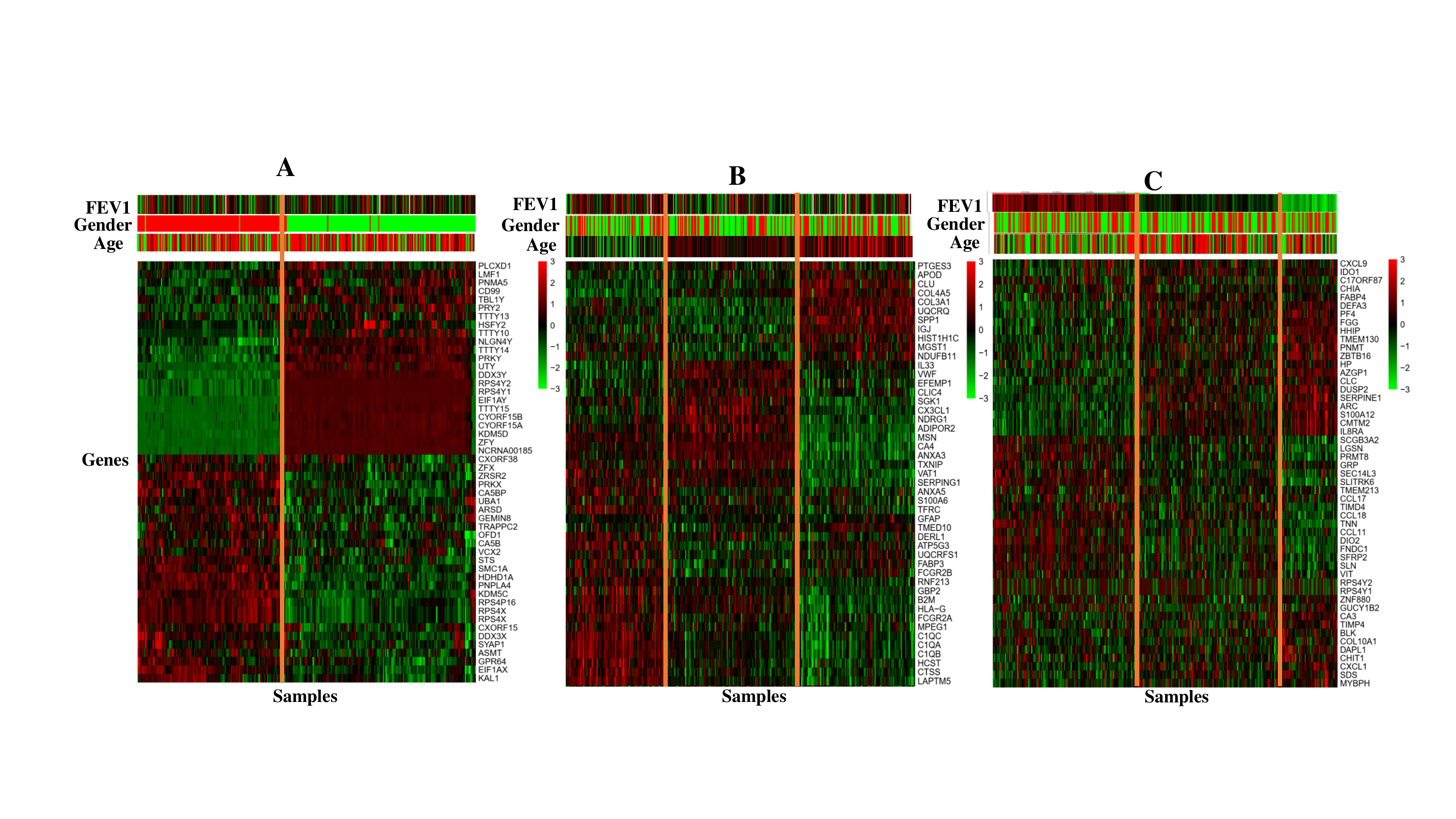}}
	\caption{A real example illustrates (A) two gender-associated clusters are found by the top 50 X/Y chromosome genes and $K$-means; (B) three age-associated clusters are detected by the top 50 age-related genes and $K$-means; (C) three clusters are identified from our algorithm, which are associated with clinical outcome FEV1 but neither associated with gender nor age. }
    \label{fig:heatmap}
\end{figure}

Although the aforementioned methods are powerful to simultaneously identify clusters and relevant features, the resulting clusters of patients may not guarantee biological meaning or clinical impact. A common practice is to perform post-hoc analyses to assess association between the identified clusters and disease relevant measures or clinical outcomes, such as survival. Such association justifies potential clinical relevance of the novel disease subtypes and supports further investigation. However, if no association is observed, the cluster analysis is considered a failed effort to bring clinical impact. In the clustering of high-dimensional omics data, the latter situation happens frequently since decision of final clusters largely depends on the selected features. The data may contain multi-faceted cluster structures that can be defined by different sets of gene features. In Figure \ref{fig:heatmap}, we demonstrate this phenomenon using a lung disease transcriptomic dataset. When we select the top 50 X/Y chromosome genes (annotated in the GeneCards database; www.genecards.org) that are most associated with the gender variable and perform simple $K$-means, Figure \ref{fig:heatmap}A identifies two clear male/female clusters. Similarly, if the top 50 genes associated with the age variable are selected from age-related genes annotated in the HAGR database \citep{tacutu2018human}, Figure \ref{fig:heatmap}B finds three clusters of young, middle-aged and old patients through $K$-means clustering. Although heatmaps in Figures \ref{fig:heatmap}A and \ref{fig:heatmap}B show well-separated clusters, they are not novel for the clinical purpose of disease subtyping. Figure \ref{fig:heatmap}C shows result of the proposed outcome-guided clustering method to be introduced. With guidance from the clinical outcome FEV1 (measuring the volume of air a person can exhale during the first second of forced expiration), three clusters of patients are identified with distinct clinical behavior and molecular mechanisms (see Section 4 for detailed result). When gene signals are largely driven by potentially disease-irrelevant factors (e.g., as in Figures \ref{fig:heatmap}A and \ref{fig:heatmap}B), genes that are directly relevant to the disease with greater clinical potential (e.g. Figure \ref{fig:heatmap}C) are less likely to be uncovered. In the literature, constraints in the forms of prior knowledge in samples \citep{wagstaff2001constrained} or pathway structure in features \citep{huo2017integrative} have been used to restrict the free parameters in high-dimensional space during clustering. The approaches improve biological relevance of the finding, but still cannot prevent the true outcome-associated disease subtypes from being masked by  disease-irrelevant clusters.

This practical example raises a fundamental question in clustering of high-dimensional omics data for disease subtyping: can we simultaneously identify disease subtypes and the driving gene signatures, where the detection of disease subtypes is guided by outcome association? This question is unique as it touches both supervised and unsupervised components in the context of machine learning. In the process of detecting novel disease subtypes, we focus on identifying disease-related subtypes and hope to disentangle and reduce impact of factors driven by clinically irrelevant variables (e.g. demographic variables, such as gender, age and race). In the literature, little has been done in this proposed direction. \cite{bair2004semi} and \cite{koestler2010semi} have developed a two-stage semi-supervised method, where  $K$-means or other standard clustering methods are applied to the top $M$ features with the highest marginal outcome association. The two-stage approach is, however, ad hoc in selecting the number of top features $M$ and has difficulty in incorporating confounding variables in the outcome association. In this paper, we propose an outcome-guided clustering (ogClust) model to provide a unified solution. To the best of our knowledge, the proposed method is the first unified generative model for outcome-guided disease subtyping (clustering).

Throughout this paper, we avoid the term ``semi-supervised'' adopted by \cite{bair2004semi} and \cite{koestler2010semi}. Instead, we name by ``outcome-guided disease subtyping’’ or ``outcome-guided clustering’’ since the term ``semi-supervised learning’’ has been used in at least two other machine learning scenarios: (1) A small set of labeled data and a larger set of unlabeled data are jointly analyzed for machine learning; (2) Cluster analysis is pursued with known constraints (e.g. pairs of observation must or must not be clustered together). Interested readers may refer to \cite{bair2013semi} for a review of semi-supervised clustering methods. One should also note that the outcome-guided clustering discussed in this paper substantially differs from latent class analysis methods in regression setting by \cite{houseman2006feature}, \cite{desantis2007penalized} and \cite{desantis2012supervised}. In this case, patients in latent classes are identified to have heterogeneous intercepts or regression slopes. The latent classes, in a sense, represent patient clusters (or disease subtypes), but there is lack of a gene signature and prediction model to classify future patients into the disease subtypes (latent classes), presenting a major obstacle towards precision medicine.

The paper is structured as follows. Section \ref{s:model} introduces the ogClust model (Section 2.1, an EM algorithm for parameter estimation (Section 2.2), extensions to robust estimation procedures in outcome association (Section 2.3), and its extension to survival outcome (Section \ref{ss:survival}). We perform extensive simulations to evaluate ogClust and compare it with existing methods in Section \ref{s:sim}, and evaluate its robust estimation in Section \ref{s:robust_sim}. A disease subtyping application using a lung disease transcriptomic dataset is presented in Section \ref{s:real}. We include final conclusion and discussion in Section \ref{s:conclusion}.

\section{Outcome-guided Clustering Model}
\label{s:model}
\subsection{Model and Notations}
We consider the problem of disease subtyping (clustering) of $n$ observations from high-dimensional data $\mathbb{G}=\{g_{ij}, 1\leq i \leq n, 1 \leq j \leq q \}$, where $\mathbb{G}$ can be mRNA expression, miRNA expression, methylation or phenomic data and $q$ can be at the scale of hundreds to thousands. Our ultimate goal is to cluster $n$ observations into $K$ clinically meaningful clusters represented by latent group label $\mathbb{Z}=\{ z_i, 1 \leq i \leq n \}$, $z_i \in \{ 1,\dots,K \}$, and $ z_i = k$ means that observation $i$ is assigned to cluster $k$ $(1 \leq k \leq K)$. Since clustering result purely from $\mathbb{G}$ may not necessarily be clinically useful as discussed in Section \ref{s:intro}, we assume that a clinical outcome $\mathbb{Y}= \{y_i, 1\leq i \leq n \}$ is given to guide the clustering (e.g. survival outcome or FEV1\%prd in the lung disease example in Section \ref{s:real}). We also assume a set of pre-specified covariates $\mathbb{X}= \{ x_{ij}, 1 \leq i \leq n, 1 \leq j \leq p \}$, where the $p$ covariates (e.g. age, gender, etc.) are potentially associated with the outcome and may confound with the association between ${Z}$ and ${Y}$. Denote by $\boldsymbol{g}_i = (g_{i1}, \dots, g_{iq})^T$ and $\boldsymbol{x}_i= (x_{i1}, \dots, x_{ip})^T$. We assume observed data $(y_i, \boldsymbol{x}_i, \boldsymbol{g}_i)$ for subject $i$ $(1\leq i \leq n)$ are independent realizations of the model for $(Y, \boldsymbol{X}, \boldsymbol{G})$.

\begin{figure}[!b]
	\noindent\makebox[\textwidth]{\includegraphics[scale=0.4]{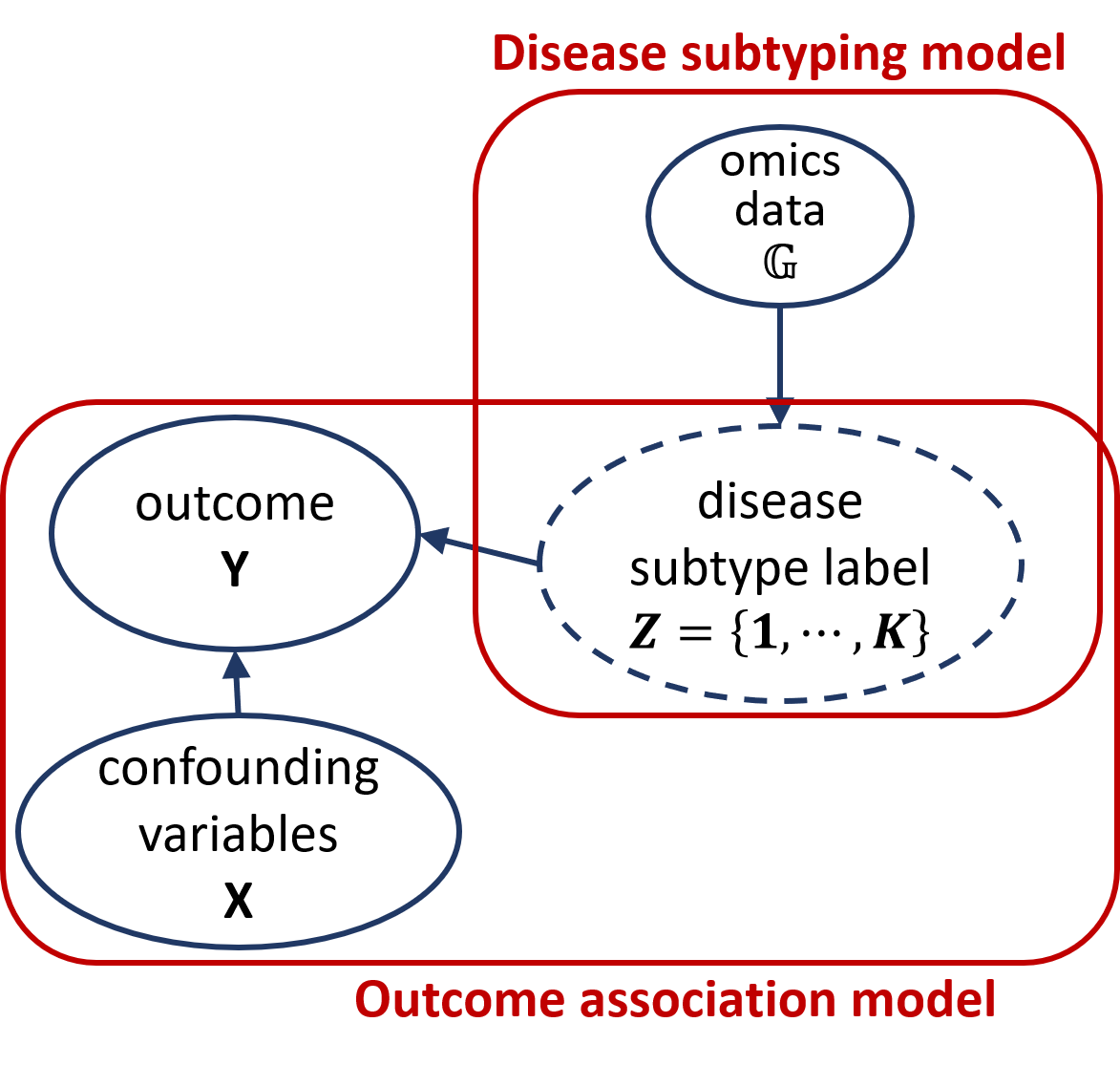}}
     \caption{A graphical illustration of the unified regression model. Y is the outcome to guide clustering, X are the baseline covariates that are believed to have effects on Y. G are the variables (e.g. gene expression) that defines the outcome associated subgroups. Z is the unobserved latent subgroup index to define final clustering.}
    \label{fig:model}
\end{figure}

As shown in Figure \ref{fig:model}, the proposed ogClust framework consists of two components: \textit{disease subtyping model} and \textit{outcome association model}. The disease subtyping model is a conventional high-dimensional discriminant analysis where we train to characterize $\pi_k = Pr(Z=k|\boldsymbol{G})$ (or $\pi_{i k} = Pr(Z_i=k|\boldsymbol{g}_i)$ for observation $i$). In this paper, we apply a multinomial logistic regression $
\pi_{ik}|\boldsymbol{\gamma} = \frac{\exp \left(\boldsymbol{g}_{i}^{T} \boldsymbol{\gamma}_{k}\right)}{\sum_{l=1}^{K} \exp \left(\boldsymbol{g}_{i}^{T} \boldsymbol{\gamma}_{l}\right)}$, where $\boldsymbol{\gamma}=\{\boldsymbol{\gamma}_k, 1 \leq k \leq K\}$
and $\boldsymbol{\gamma}_{k} = (\gamma_{1k}, \dots, \gamma_{qk})^T$. Since $q$ is usually large, we assume only a small subset ${\mathcal{A}} \subset \{1, \dots, q\}$ of features effective in characterizing the clusters that affect the outcome, where its cardinality $\operatorname{card}(\mathcal{A}) < \min(n, q)$. In other words, $\boldsymbol{\gamma}_{[j]} \neq \boldsymbol{0}$ if $j \in \mathcal{A}$ and $\boldsymbol{\gamma}_{[j]} = \boldsymbol{0}$ if $j \in \mathcal{A}^c$, where $\boldsymbol{\gamma}_{[j]}=\{ \gamma_{j1}, \dots, \gamma_{jK}\}$. We apply LASSO regularization, or group LASSO regularization \citep{tibshirani2012strong} with parameters in $\boldsymbol{\gamma}_{[j]}$ as a group to the multinomial logistic regression to select subtyping features.

In the outcome association model, we assume the following mixture model:
\begin{equation}
    f(y_i ; \boldsymbol{x_i})=\sum_{k=1}^{K} \pi_{ik} f_k (y_i ; \boldsymbol{x_i} ),
    \label{equation1}
\end{equation}
where $f_k(y; \boldsymbol{x})$ is density function of cluster $k$. We assume a continuous response $Y$ where the $k$-th mixture density $f_k(y; \boldsymbol{x}, \beta_{0k}, \boldsymbol{\beta}, \sigma)$ is parameterized by cluster specific intercept $\beta_{0k}$, common covariate effect $\boldsymbol{\beta} = (\beta_{1}, \dots, \beta_{p})^T$ and a homogeneous error $\sigma$. In this paper, we specifically assume $y_i|z_i=k \sim N(\beta_{0k} + \boldsymbol{\beta}^T \boldsymbol{x_i}, \sigma^2)$ with mixture probability $\pi_{ik}=\frac{\exp \left(\boldsymbol{g}_{i}^{T} \boldsymbol{\gamma}_{k}\right)}{\sum_{l=1}^{K} \exp \left(\boldsymbol{g}_{i}^{T} \boldsymbol{\gamma}_{l}\right)}$, $k = 1,\dots, K$. Denote by
$\boldsymbol{\theta} = \{\boldsymbol{\beta}_0, \boldsymbol{\beta}, \boldsymbol{\gamma},  \sigma \}$ the collection of all parameters from the two models in ogClust ($\boldsymbol{\beta}_0 = (\beta_{01}, \dots, \beta_{0K})^T$), given $\mathbb{Y}$, $\mathbb{X}$ and $\mathbb{G}$, $\boldsymbol{\theta}$ can be estimated by maximizing the following sample likelihood of the basic model:
\begin{equation}
    \mathrm{L}(\boldsymbol{\theta})=\prod_{i=1}^{n} \sum_{k=1}^{K} \pi_{i k}(\boldsymbol{g_i}, \boldsymbol{\gamma}) f\left(y_{i} ; \boldsymbol{x}_{i}, \beta_{0 k}, \boldsymbol{\beta}, \sigma  \right).
    \label{equation2}
\end{equation}

\noindent \textit{Remarks:}
\begin{enumerate}[leftmargin=*]
    \item Generalization from continuous outcome $Y$ to other types of outcome $Y$ is relatively straightforward. Section \ref{ss:survival} discusses the extension to survival outcome.
    \item In the current model, we assume only several important and pre-selected covariates for $\boldsymbol{X}$ and no variable selection is implemented in the outcome association model. Including $\boldsymbol{X}$ (e.g. age or gender) in outcome association has two main advantages: (i) it corrects for potential confounding effects between the association of outcome $Y$ and subtype $Z$, (ii) if a covariate, say, gender, is indeed predictive of $Y$ and there exist many strong gender-associated genes in $\boldsymbol{G}$, the model will avoid identification of gender-related clusters in $Z$. In this case, although gender-associated subtypes are predictive of the outcome, their information has been captured by observable covariate and thus can be avoided in subtyping.
    \item The current model assumes a simplified common covariate effect $\boldsymbol{\beta}$ across all clusters. It is straightforward to extend for cluster-specific interaction term $\boldsymbol{\beta_k}$, meaning cluster-specific age or gender effects.
    \item We apply multinomial logistic regression in this paper but other high-dimensional discriminant analysis methods, such as sparse linear discriminant analysis, can also be used.
    \item The conditional probability $\hat{\pi}_{ik}|\boldsymbol{\hat{\gamma}} = \frac{\exp \left(\boldsymbol{g}_{i}^{T} \boldsymbol{\hat{\gamma}}_{k}\right)}{\sum_{l=1}^{K} \exp \left(\boldsymbol{g}_{i}^{T} \boldsymbol{\hat{\gamma}}_{l}\right)}$ can be used to predict the cluster label of new observations.
\end{enumerate}

\subsection{Numerical Solution with Gene selection}
A numerical method using EM algorithm is proposed for ogClust parameter estimation in Eq. (2). By introducing $z_{ik}, k =1,\dots, K$, as missing indicator variables, following the seminal idea in \cite{dempster1977maximum}, the complete log likelihood function can be written as  
\begin{equation}
l_{n}^{c}(\boldsymbol{\theta})=\sum_{i=1}^{n} \sum_{k=1}^{K}\left\{z_{i k} \log \pi_{i k}+z_{i k} \log f\left(y_{i} ; \boldsymbol{x}_{i}, \beta_{0k}, \boldsymbol{\beta}, \sigma \right)\right\},
\label{equation3}
\end{equation}
where $z_{ik}=1$ if subject $i$ belongs to subgroup $k$, and $z_{ik}=0$ otherwise.

Since gene expression is usually high dimensional, including genes in $\mathcal{A}^c$ with non-predictive effect will introduce extra noise to the disease subtyping model and may produce irrelevant subtypes that are not necessarily related to the disease outcome of interest. In the following, we will illustrate with a LASSO penalty or an alternative group LASSO regularization framework for gene selection. We define the penalized log-likelihood function as
\begin{equation}
\tilde{l}_{n}^{c}(\boldsymbol{\theta})=\sum_{i=1}^{n} \sum_{k=1}^{K}\left\{z_{i k} \log \pi_{i k}+z_{i k} \log f\left(y_{i} ; \boldsymbol{x}_{i}, \beta_{0k}, \boldsymbol{\beta},  \sigma \right)\right\}-\lambda R(\boldsymbol{\gamma}),
\label{equation4}
\end{equation}
where $\lambda$ is the regularization tuning parameter and $R(\boldsymbol{\gamma})=\sum_{j=1}^{q} \sum_{k=1}^{K}|\gamma_{j k}|$ for LASSO penalty. Alternatively, we can use group LASSO penalty plus $\ell_2$ regularization $R(\boldsymbol{\gamma})=\sum_{j=1}^{q}\|\boldsymbol{\gamma}_{[j]}\|_{2} + \alpha\sum_{j=1}^{q}\sum_{k=1}^{K}\gamma_{j k}^2$, where $\|\boldsymbol{\gamma_{[j]}}\|_{2}=\sqrt{\sum_{k=1}^K \gamma_{j k}^2}$.  The first term is a group LASSO penalty to select or deselect $\boldsymbol{\gamma_{[j]}}$ for gene $j$. The second term encourages joint selection of predictive genes with high collinearity. Detecting multiple genes with high collinearity offers better molecular insight to the subtype mechanism and provides more stable cluster prediction for future patients. The irrelevant features are removed by shrinking corresponding elements of $\boldsymbol{\gamma_{[j]}}$ to zero, thus a sub-model is automatically selected. This procedure performs feature selection and numerical estimation of parameters simultaneously. 

Maximization of $\tilde{l}_{n}^{c}(\boldsymbol{\theta})$ can be achieved by sequentially and iteratively updating $\boldsymbol{\beta_0}$, $\boldsymbol{\beta}$, $\sigma$ and $\boldsymbol{\gamma}$ in an EM algorithm, which takes the following steps:
\begin{itemize}
    \item The E step computes the conditional expectation of the function $\tilde{l}_{n}^{c}(\boldsymbol{\theta})$ with respect to $z_{ik}$, given the observed data $y_i$, $\boldsymbol{x}_i$ and the current parameter estimates $\boldsymbol{\theta}^{\left( m \right)}$, 
    $$
    Q\left(\boldsymbol{\theta}, \boldsymbol{\theta}^{(m)}\right)=\sum_{i=1}^{n} \sum_{k=1}^{K} w_{i k}^{(m)} \log \pi_{i k}+\sum_{i=1}^{n} \sum_{k=1}^{K} w_{i k}^{(m)} \log f\left(y_{i} ; \boldsymbol{x}_{i}, \beta_{0k}, \boldsymbol{\beta}, \sigma \right)-\lambda \sum_{j=1}^{q}R(\boldsymbol{\gamma}_j),
    $$
    where the posterior weights 
    \begin{equation}
    w_{i k}^{(m)}=E\left(Z_{i k} | y_{i}, \boldsymbol{x}_{i}, \boldsymbol{\theta}^{(m)}\right)=\frac{\pi_{i k}^{(m)} f\left(y_{i} ; \boldsymbol{x}_{i}, \beta_{0k}^{(m)}, \boldsymbol{\beta}^{(m)},  \sigma^{(m)}\right)}{\sum_{l=1}^{K} \pi_{i l}^{(m)} f\left(y_{i} ; \boldsymbol{x}_{i}, \beta_{0l}^{(m)}, \boldsymbol{\beta}^{(m)}, \sigma^{(m)}\right)}.
    \label{equation5}
    \end{equation}
\item The M step on the $(m+1)$-th iteration maximizes the $Q\left(\boldsymbol{\theta}, \boldsymbol{\theta}^{(m)}\right)$ with respect to $\boldsymbol{\theta}$. By taking partial derivative, it is easy to show that
    $\boldsymbol{\beta}_{0}$, $\boldsymbol{\beta}$ and $\sigma^2$ are updated by the following updating equations:
    \begin{eqnarray}
        \beta_{0 k}^{(m+1)} & = & \frac{\sum_{i=1}^{n} w_{i k}^{(m)}\left(y_{i}-(\boldsymbol{\beta}^{(m)})^T\boldsymbol{x}_{i}         \right)}{\sum_{i=1}^{n} w_{i k}^{(m)}}, \quad
        k = 1, \dots, K,
        \label{equation6} \\
        \beta_{\ell}^{(m+1)} & = & \frac{\sum_{i=1}^{n} \sum_{k=1}^{K} w_{i k}^{(m)} x_{i \ell}\left(y_{i}-\beta_{0k}^{(m+1)}-\sum_{h \neq \ell} \beta_{h }^{(m)} x_{i h}\right)}{\sum_{i=1}^{n} \sum_{k=1}^{K} w_{i k}^{(m)} x_{i \ell}^{2}}, \quad
        \ell = 1, \dots, p,
        \label{equation7} \\
    (\sigma^{(m+1)})^2 & = & \frac{\sum_{i=1}^{n} \sum_{k=1}^{K} w_{i k}^{(m)}\left(y_{i}-\beta_{0k}^{(m+1)}- (\boldsymbol{\beta}^{(m+1)})^T \boldsymbol{x}_{i} \right)^{2}}{\sum_{i=1}^{n} \sum_{k=1}^{K} w_{i k}^{(m)}}.
            \label{equation8}
    \end{eqnarray}
    The updated estimates $\boldsymbol{\gamma}^{(m+1)}$ is obtained following an approximation procedure of \cite{friedman2010regularization}. For lasso penalty $R(\boldsymbol{\gamma})=\sum_{k=1}^{K} R_k(\boldsymbol{\gamma}_k)= \sum_{k=1}^{K} \sum_{j=1}^{q}  |\gamma_{j k}|$, the likelihood for estimating $\boldsymbol{\gamma}^{(m+1)}$ given $w^{(m)}$ is 
    \begin{eqnarray}
        \begin{aligned} \tilde{l}_p\left(\boldsymbol{\theta}, \boldsymbol{\theta}^{(m)}\right)
        &= \sum_{i=1}^{n} \sum_{k=1}^{K} w_{i k}^{(m)} \log \pi_{i k} -\lambda R(\boldsymbol{\gamma}) =\sum_{i=1}^{n} \sum_{k=1}^{K} w_{i k}^{(m)} \log \frac{\exp \left(\boldsymbol{g}_{i}^{T} \boldsymbol{\gamma}_{k}\right)}{\sum_{l=1}^{K} \exp \left(\boldsymbol{g}_{i}^{T} \boldsymbol{\gamma}_{l}\right)} -\lambda R(\boldsymbol{\gamma}) \\ &= \sum_{i=1}^{n}\sum_{k=1}^{K} w_{i k}^{(m)} \left\{ \boldsymbol{g}_{i}^{T} \boldsymbol{\gamma}_{k}-\log \left(\sum_{l=1}^{K} \exp \left(\boldsymbol{g}_{i}^{T} \boldsymbol{\gamma}_{l}\right)\right)\right\}  -\lambda R(\boldsymbol{\gamma})
        \end{aligned}
        \label{equation9}
    \end{eqnarray}
We approximate the partial log likelihood $\tilde{l}_p(\boldsymbol{\theta}, \boldsymbol{\theta}^{(m)})$ by quadratic approximation. The resulting partial likelihood $\tilde{l}_{Q k}(\boldsymbol{\theta}, \boldsymbol{\theta}^{(m)})$ for subgroup $k$ is in the form of a weighted least square: 
$$
\tilde{l}_{Q k}\left(\boldsymbol{\theta}, \boldsymbol{\theta}^{(m)}\right)=-\frac{1}{2} \sum_{i=1}^{n} W_{i k}\left(h_{i k}-\boldsymbol{g}_{i}^{T} \boldsymbol{\gamma}_{k}\right)^{2}-\lambda R_k(\boldsymbol{\gamma}_k) + C,
$$
where
${h_{i k}=\boldsymbol{g}_{i}^{T} \boldsymbol{\gamma}_{k}^{(m)}+\frac{w_{i k}^{(m)}-\pi_{i k}^{(m)}}{W_{i k}}}$, ${W_{i k}=\pi_{i k}^{(m)}(1-\pi_{i k}^{(m)})}$, and $C$ is independent of $\boldsymbol{\gamma}_k$.
Thus the solution to $\boldsymbol{\gamma}^{(m+1)}$ can be obtained by coordinate descent, i.e., individually solving 
$\max_{\boldsymbol{\gamma}_{k} \in R^{q}} \tilde{l}_{Q k}(\boldsymbol{\theta}, \boldsymbol{\theta}^{(m)}) $ for each $k$. By some algebraic manipulation, we obtain the  estimate 
    \begin{equation}
    \tilde{\gamma}_{k j}=\frac{S\left(\sum_{i=1}^{N} g_{i j} W_{i k}\left(h_{i k}-(\boldsymbol{g}_{i}^{(j)})^{T} \tilde{\boldsymbol{\gamma}}_{k}^{(j)}\right), \lambda\right)}{\sum_{i=1}^{N} W_{i k} g_{i j}^{2}},
        \label{equation10}
    \end{equation}
    where 
    $ S(z, \lambda) = \operatorname{sign}(z)(|z|-\lambda)_{+} 
    $ is a soft thresholding operator, $(a)_+ = \max(0, a)$, $\tilde{\boldsymbol{\gamma}}_{k}^{(j)}$ is the parameter vector $\tilde{\boldsymbol{\gamma}}_{k}$ omitting $\tilde{\gamma}_{k j}$, and $\boldsymbol{g}_{i}^{(j)}$ is the gene vector $\boldsymbol{g}_{i}$ omitting $g_{ij}$. The coordinate descent procedure iteratively updates the current estimate $\boldsymbol{\tilde{\gamma}}$ until convergence. For the group LASSO + $\ell_2$ regularization, we apply the glmnet function in R package glmnet, setting multinomial family, grouped type and $\alpha$ equals 0.5.
\end{itemize}

\begin{algorithm}[t]
\SetAlgoLined
\SetKwInOut{Input}{input}
\SetKwInOut{Output}{output}
 \vspace{3pt}
\textbf{input:} $\mathbb{Y}$, $\mathbb{X}$, $\mathbb{G}$ and $K$ \\
 Initialize $\boldsymbol{\theta}^{(0)}$ and set $m = 0$;\\
 \Repeat{$||\boldsymbol{\theta}^{(m)}-\boldsymbol{\theta}^{(m-1)}||< 10^{-7}$}{
  E-step: 
  compute the posterior weights $w_{i k}^{(m)}$ by Equation \eqref{equation5};\\
  M-step: \\
  1. Update $\{\boldsymbol{\beta}_0^{(m)}, \boldsymbol{\beta}^{(m)}, \sigma^{(m)}\}$ to $\{\boldsymbol{\beta}_0^{(m+1)}, \boldsymbol{\beta}^{(m+1)}, \sigma^{(m+1)} \}$ by Equations \eqref{equation6}-\eqref{equation8};\\
  2. Update $\boldsymbol{\gamma}^{(m)}$ to $\boldsymbol{\gamma}^{(m+1)}$ by coordinate descent: \\
  Set $\tilde{\boldsymbol{\gamma}}^{old} = \boldsymbol{\gamma}^{(m)}$; \\
  \Repeat{$||\tilde{\boldsymbol{\gamma}}^{old} - \tilde{\boldsymbol{\gamma}}^{new}|| < 10^{-7}$}{
   Update $\tilde{\gamma}_{k j}^{old}$ to $\tilde{\gamma}_{k j}^{new}$ by Equation \eqref{equation10}, for $k = 1,\dots, K$ and $j=1,\dots,q$;
   }
   Set $\boldsymbol{\gamma}^{(m+1)} = \tilde{\gamma}_{k j}^{new}$, $\boldsymbol{\theta}^{(m+1)}$ = $\{ \boldsymbol{\beta}_0^{(m+1)}, \boldsymbol{\beta}^{(m+1)}, \boldsymbol{\gamma}^{(m+1)}, \sigma^{(m+1)} \}$, 
   $m = m + 1$;
 }
 \Output{Parameter estimates $ \boldsymbol{\hat{\theta}} = \boldsymbol{\theta}^{(m)}$}
 \vspace{3pt}
\caption{Pseudo code for ogClust model estimation.}
\label{algorithm1}
\end{algorithm}

The pseudo code for fitting the unified ogClust model is given in Algorithm~\ref{algorithm1}. Multiple initials could be used to avoid convergence to local minimums and increase the numerical stability of parameter estimates. We use Bayesian information criterion (BIC) to determine the tuning parameter $\lambda$ and the number of subgroups $K$ in simulation. BIC is defined as $\ln(n)\mbox{df}(\hat{\boldsymbol{\theta}}) - 2\ln(\mbox{L}(\hat{\boldsymbol{\theta}}))$, where $\boldsymbol{\hat{\theta}} = \{\boldsymbol{\hat{\beta}}_0, \boldsymbol{\hat{\beta}}, \boldsymbol{\hat{\gamma}},  \hat{\sigma} \}$ and df($\hat{\boldsymbol{\theta}}$) is the number of non-zero estimated parameters. In the real application, because of potential data noises and violation of Gaussian assumption, BIC may fail to choose the correct $K$. To address this issue, we plot the trend of RMSE and $R^2$ as a function of $K$ and identify the elbow point as the optimal number of clusters $K$ as shown in Figure S3. 

\subsection{Robust Estimation Procedures}
The ogClust model is based on and could be sensitive to the Gaussian mixture assumption in outcome $Y$. There are three common types of model misspecification: (A) heavy-tailed or skewed error term in the outcome association model, (B) outlier outcomes in the outcome association model, and (C) scattered observations who do not fit into any of the $K$ subtypes in the disease subtyping model. Our model is relatively robust to type C noises because of the soft assignment using multinomial logistic probability function. One may iteratively remove a small number of samples with unconfident cluster assignment. To guard against the first two types of model misspecification, we propose 1) ogClust with median-truncated loss (ogClust-median-truncation) 2) ogClust with Huber loss (ogClust-Huber) 3) ogClust with adaptive-Huber loss (ogClust-adHuber) to replace the original ogClust with quadratic loss. Intuitively, median-truncation and Huber loss functions are effective in dealing with potential outliers.  As we will introduce later, the adaptive-Huber loss is particularly useful for heavy-tailed and skewed error terms.
Hence, the penalized log-likelihood function is defined as
$$l_{n}^{c}(\theta)=\sum_{i=1}^{n} \sum_{k=1}^{K}\left\{z_{i k} \log \pi_{i k}+z_{i k} \ell_{\tau}(e_{ik})\right\}-\lambda \sum_{j=1}^{p}R(\boldsymbol{\gamma}_j).$$
where $\ell_{\tau}(e_{ik})$ denotes the robust loss function to replace $\log f\left(y_{i} ; \boldsymbol{x}_{i}, \beta_{0k}, \boldsymbol{\beta},  \sigma \right)$. We follow the same EM procedure with modified loss functions to compute numerical solutions.

\subsubsection{\underline{Median-truncated Loss}} 
The median-truncated loss \citep{chi2019median} describes the loss function for subject $i$ in subgroup $k$ as:
        $$
        \ell_{\tau}(e_{ik})=\left\{\begin{array}{ll}{e_{ik}^{2} / 2} & {\text { if }|e_{ik}| \leq \tau_k} \\ {0} & {\text { if }|e_{ik}|>\tau_k}\end{array}\right.,
        $$
        where $e_{ik}=y_i-\hat{\beta}_{0 k}- \hat{\boldsymbol{\beta}}^T \boldsymbol{X}_{i} $, and $\tau_k= \text{median}\left\{ |e_{ik}| \right\}_{i=1}^n$.
        The loss function remains the same for $e_{ik}$ smaller or equal to median $\tau_k$, and the loss function equals to 0 for $e_{ik}$ larger than the median $\tau_k$. The cutoff $\tau_k$ is chosen as the median of $e_{1k}, ..., e_{nk}$. By taking partial derivative, the estimates in the $(m+1)$th iteration for $\boldsymbol{\beta}_{0}$, $\boldsymbol{\beta}$ and $\sigma$ are obtained by the following equations:
\begin{eqnarray}
        \beta_{0 k}^{(m+1)} &=& \frac{\sum_{i=1}^{n} w_{i k}^{(m)}\left(y_{i}-(\boldsymbol{\beta}^{(m)})^T \boldsymbol{x}_{i}  \right) I\left(\left|e_{ik}^{(m)}\right|\leq\tau_k\right)}{\sum_{i=1}^{n} w_{i k}^{(m)} I\left(\left|e_{ik}^{(m)}\right|\leq\tau_k\right)}, \nonumber \\
        \beta_{\ell}^{(m+1)} &=& \frac{\sum_{i=1}^{n} \sum_{k=1}^{K} w_{i k}^{(m)} x_{i \ell}\left(y_{i}-\beta_{0 k}^{(m+1)}-\sum_{h \neq \ell} \beta_{h}^{(m)} x_{i h}\right)I\left(\left|e_{ik}^{(m)}\right|\leq\tau_k\right)}{\sum_{i=1}^{n} \sum_{k=1}^{K} w_{i k}^{(m)} x_{i\ell}^{2}I\left(\left|e_{ik}^{(m)}\right|\leq\tau_k\right)}, \nonumber \\
        (\sigma^{(m+1)})^2 &=& \frac{\sum_{i=1}^{n} \sum_{k=1}^{K} w_{i k}^{(m)}\left(y_{i}-\beta_{0 k}^{(m+1)}- (\boldsymbol{\beta}^{(m+1)})^T \boldsymbol{x}_{i} \right)^{2}I\left(\left|e_{ik}^{(m)}\right|\leq\tau_k\right)}{\sum_{i=1}^{n} \sum_{k=1}^{K} w_{i k}^{(m)}I\left(\left|e_{ik}^{(m)}\right|\leq\tau_k\right)}. \nonumber
\end{eqnarray}
        
\subsubsection{\underline{Huber Loss}}
The Huber loss alternatively describes the loss function for subject $i$ in subgroup $k$ as:
        $$
        \ell_{\tau}(e_{ik})=\left\{\begin{array}{ll}{e_{ik}^{2} / 2} & {\text { if }|e_{ik}| \leq \tau} \\ {\tau|e_{ik}|-\tau^{2} / 2} & {\text { if }|e_{ik}|>\tau}\end{array}\right..
        $$
 This loss function is quadratic for small values of $e$, and linear for large values of $e$. The cutoff $\tau$ is suggested as a fixed constant ($\tau=1.345$) which gives 95\% efficiency under Gaussian assumption in regression setting \citep{huber2004robust}. By EM algorithm, the estimates in the $(m+1)$th iteration for $\boldsymbol{\beta}_{0}$, $\boldsymbol{\beta}$ and $\sigma^2$ are obtained by the following equations:
\begin{eqnarray}
        \beta_{0 k}^{(m+1)} &=& \frac{\sum_{i=1}^{n} w_{i k}^{(m)}\left(y_{i}- (\boldsymbol{\beta}^{(m)})^T \boldsymbol{x}_{i}  \right) I\left(\left|e_{ik}^{(m)}\right|\leq\tau\right)+\sum_{i=1}^{n} w_{i k}^{(m)} \cdot \tau \cdot \operatorname{sign}\left(e_{i k}^{(m)}\right) \cdot I\left(\left|e_{ik}^{(m)}\right|>\tau\right)}{\sum_{i=1}^{n} w_{i k}^{(m)} I\left(\left|e_{ik}^{(m)}\right|\leq\tau\right)}, \nonumber\\
        \beta_{\ell}^{(m+1)} &=& \frac{\sum_{i=1}^{n} \sum_{k=1}^{K} w_{i k}^{(m)} x_{i \ell}\left(\left(y_{i}-\beta_{0 k}^{(m+1)}-\sum_{h \neq \ell} \beta_{h}^{(m)} x_{i h}\right)I\left(\left|e_{i k}^{(m)}\right|\leq\tau\right)\right)}{\sum_{i=1}^{n} \sum_{k=1}^{K} w_{i k}^{(m)} x_{i \ell}^{2}I\left(\left|e_{i k}^{(m)}\right|\leq\tau\right)}
        + \nonumber\\
        && \frac{\tau\operatorname{sign}\left(e_{i k}^{(m)}\right)I\left(\left|e_{i k}^{(m)}\right|>\tau\right)}{\sum_{i=1}^{n} \sum_{k=1}^{K} w_{i k}^{(m)} x_{i \ell}^{2}I\left(\left|e_{i k}^{(m)}\right|\leq\tau\right)},\nonumber\\
        (\sigma^{(m+1)})^2 &=& \frac{\sum_{i=1}^{n} \sum_{k=1}^{K} w_{i k}^{(m)}\left(y_{i}-\beta_{0 k}^{(m+1)}- (\boldsymbol{\beta}^{(m+1)})^T \boldsymbol{x}_{i} \right)^{2}I\left(\left|e_{ik}^{(m)}\right|\leq\tau\right)}{\sum_{i=1}^{n} \sum_{k=1}^{K} w_{i k}^{(m)}I\left(\left|e_{ik}^{(m)}\right|\leq\tau\right)} + \nonumber\\
        &&\frac{\sum_{i=1}^{n} \sum_{k=1}^{K} w_{i k}^{(m)}\left(2\tau\left|y_{i}-\beta_{0 k}^{(m+1)}-\boldsymbol{X}_{i}^{T} \boldsymbol{\beta}^{(m+1)}\right|-\tau^2\right)I\left(\left|e_{ik}^{(m)}\right|>\tau\right)}{\sum_{i=1}^{n} \sum_{k=1}^{K} w_{i k}^{(m)}I\left(\left|e_{ik}^{(m)}\right|\leq\tau\right)}. \nonumber
\end{eqnarray}

\subsubsection{\underline{Adaptive Huber Loss}}
When there is no outlier but the error term is heavy-tailed asymmetric, median-truncated loss or Huber loss using constant $\tau$ would introduces bias \citep{sun2019adaptive}. To mitigate this bias, we use an adaptive Huber loss in the EM algorithm by adopting the method of \cite{wang2020}. In this method, the cutoff $\tau$ is data-driven and estimated adaptively, taking into account sample size, $n$, dimension of $\boldsymbol{\beta}$, $p$, by iteratively solving the following equations:
$$\left\{\begin{array}{l}
g_{1}(\boldsymbol{\theta}, \tau):=\sum_{i=1}^{n} w_{ik}^{(m)}\sum_{k=1}^{K} \ell^{'}_{\tau}\left(e_{ik}\right) \boldsymbol{X}_{i}=\mathbf{0} \\
g_{2}(\boldsymbol{\theta}, \tau):=(n-p)^{-1} \sum_{i=1}^{n} \sum_{k=1}^{K} \min \left\{e_{ik}^{2}, \tau^{2}\right\} / \tau^{2}-n^{-1}(p+z)=0
\end{array}\right.$$
 ,where $z=log(n)$ by default. This method is implemented in R package tfHuber. We adapt it into the M-step of our EM algorithm to update $\{\boldsymbol{\beta}_0, \boldsymbol{\beta}, \sigma \}$. At a high level, by allowing increasing value of cutoff $\tau$ as $n$ increases, there is a trade-off between the robustness and bias. By picking an optimal $\tau$, the bias becomes negligible while the result is still robust to outliers caused by heavy-tailed noises.
 
\subsection{ogClust Model with Survival Outcome}
\label{ss:survival}
ogClust model can be extended to use survival outcome. To facilitate model fitting, we choose accelerated failure time (AFT) model with log-logistic distribution to model time-to-event data as $(log(Y)|Z=k) = \beta_{0 k} + X\beta + W\sigma$, where W $\sim$ standard logistic distribution and $\sigma$ is the standard deviation. Therefore, the likelihood of mixture model can be written as
$$
\mathrm{L}(\boldsymbol{\theta})=\prod_{i=1}^{n} \sum_{k=1}^{K} \pi_{ik} L_{ik}\left(y_{i} | \boldsymbol{x}_{i}, \beta_{0 k}, \boldsymbol{\beta}, \sigma_k, \right).
$$
Denote $\delta$ as a binary indicator of event, $\delta=1$ means event and 0 means right-censored. The likelihood function $L_{ik}\left(Y_{i} | \boldsymbol{x}_{i}, \beta_{0 k}, \boldsymbol{\beta}, \sigma \right)$ is defined as
$$L_{ik}\left(y_{i} | \boldsymbol{x}_{i}, \beta_{0 k}, \boldsymbol{\beta}, \sigma \right)=\left\{\frac{1}{\sigma} f_W\left(w_i\right)\right\}^{\delta_{i}}\left\{S_W\left(w_i\right)\right\}^{1-\delta_{i}}$$
, where
$$\begin{aligned}
&w_i=\frac{z_{i}-\beta_{0 k}-X_{i} \beta}{\sigma}\\
&S_W(w_i)=1 /\left(1+e^{w_i}\right)\\
&f_W(w_i)=e^{w_i} /\left(1+e^{w_i}\right)^{2}.
\end{aligned}$$
Therefore, the penalized log-likelihood function is defined as:
$$\tilde{l}_{n}^{c}(\theta)=\sum_{i=1}^{n} \sum_{k=1}^{K}\left\{z_{i k} \log \pi_{i k}+z_{i k} logL_{ik}\left(y_{i} | \boldsymbol{x}_{i}, \beta_{0 k}, \boldsymbol{\beta}, \sigma \right)\right\}-\lambda R(\boldsymbol{\gamma}).$$
We follow the same EM algorithm in the original model, except that the likelihood of the ATF model is maximized by implementing R package "survival".

\section{Numerical Studies}
In this section, we conduct three simulations to evaluate the performance of clustering, feature selection, and outcome prediction for ogClust, robust estimation procedures of ogClust, and its extension for survival outcome respectively. In section \ref{s:sim} we assume that the continuous outcome $Y$ follows mixture of Gaussian distribution and compare the performance of ogClust with three other methods. In section \ref{s:robust_sim} we introduce outliers or skewed and heavy-tailed errors to outcome $Y$, and compare the performance of three robust estimation procedures with the non-robust ogClust method. In section \ref{s:survival_sim} we show the advantage of ogClust over three other methods with survival outcome $Y$ to guide the clustering.
\subsection{Simulations to Evaluate OgClust}
\label{s:sim}
\underline{Simulation scheme}
\begin{enumerate}
    \item Simulate $q=1000$ genes ($\boldsymbol{G}=\{G_1,...,G_{1000}\}$), among which $G_1$ to $G_{30}$ are differentially expressed (DE) across clusters while the rest of the genes are noises and their expression values are randomly drawn from the standard normal distribution (Figure S1B). Expression levels of the 30 DE genes are randomly drawn from N(1,1) and N(0,1) to form 3 $\times$ 3 clusters as specified in Figure S1A: gene set $\boldsymbol{G}_{\mathcal{A}_1}$, $\mathcal{A}_1 = \{ 1, \dots ,15 \}$, defines three clusters associated with the outcome $Y$; gene set $\boldsymbol{G}_{\mathcal{A}_2}$, $\mathcal{A}_2 = \{ 16, \dots ,30 \}$, defines three ``clinically irrelevant clusters'' that are independent of $Y$. 
    
    \item Use parameters corresponding to $\mathcal{A}_1$, $\boldsymbol{\gamma}_{\mathcal{A}_1} = (\boldsymbol{\gamma}_{1\mathcal{A}_1},\boldsymbol{\gamma}_{2\mathcal{A}_1},\boldsymbol{\gamma}_{3\mathcal{A}_1})^T$, to represent the effect of gene expression on subtyping. For identifiability, we set $\boldsymbol{\gamma}_{3 \mathcal{A}_1} = \boldsymbol{0}$. $\boldsymbol{\gamma}_{1 \mathcal{A}_1}$ and $\boldsymbol{\gamma}_{2 \mathcal{A}_1}$ vary in different models. The active set for outcome-guided subtypes is restricted to $\mathcal{A}_1$, in other words, $\boldsymbol{\gamma}_{\mathcal{A}_1^c} = \boldsymbol{0}$. 
    
    \item Given gene expression of $\boldsymbol{G}_{\mathcal{A}_1}$ and $\boldsymbol{\gamma}_{\mathcal{A}_1}$, we obtain $\pi_{ik}=\frac{\exp (\boldsymbol{g}_{i\mathcal{A}_1}^T\boldsymbol{\gamma}_{k\mathcal{A}_1} )}{\sum_{l=1}^{3}\exp (\boldsymbol{g}_{i\mathcal{A}_1}^T\boldsymbol{\gamma}_{l\mathcal{A}_{1}})}$, $k \in \{1,2, 3\}$, which represent the probability of subject $i$ belonging to the $k$th subgroup. Therefore, subgroup indicator $Z_i$ for subject $i$ is randomly drawn from a multinomial distribution with probability $\boldsymbol{p}_i=( \pi_{i1}, \pi_{i2}, \pi_{i3} )$ 
    
    \item Sample independent covariates $X_1$ and $X_2$ are sampled from normal distributions $N(1,1)$ and $N(2,1)$ respectively. Recall that $\boldsymbol{\beta} = (\beta_1,\beta_2)^T$ is the set of regression coefficients of the two covariates and $\boldsymbol{\beta}_0 = (\beta_{01}, \beta_{02},\beta_{03})^T$ represents the baseline mean of the three subgroups. We set $\boldsymbol{\beta} = (1,1)^
   T$, and $\boldsymbol{\beta}_0$ varies according to different models.
    
    \item Given the latent subgroup index $Z_i$, the outcome for subject $i$ can be simulated by $(Y_i|Z_i = k)=\beta_{0k}+\boldsymbol{X_i^{T}\beta}+e_i$, where $e_i \sim N(0, \sigma^2)$ and we set $\sigma^2 = 1$.  
\end{enumerate}

The simulation scheme is illustrated in detail in Figure S1. Let $\boldsymbol{\beta}_0 = (1, 1+\delta, 1+2\delta)^T$ and $\boldsymbol{\gamma}_{\mathcal{A}_1}=((\operatorname{rep}(\gamma, 5), \operatorname{rep}(0, 5), \operatorname{rep}(\text{-}\gamma, 5))^T, (\operatorname{rep}(\text{-}\gamma, 5), \operatorname{rep}(0, 5), \operatorname{rep}(\gamma, 5))^T, (\operatorname{rep}(0,15))^T)$, where $\operatorname{rep}(a,b) = (a, \dots, a)_{(1 \times b)} $. We consider four models with different choices of $\boldsymbol{\beta}_0$ and $\boldsymbol{\gamma}_{\mathcal{A}_1}$ specified below: 
\begin{itemize}
    \item Model 1: $\gamma=1$ and $\delta=2$
    \item Model 2: $\gamma=1$ and $\delta=3$
    \item Model 3: $\gamma=1$ and $\delta=5$
    \item Model 4: $\gamma=3$ and $\delta=3$
\end{itemize}
Essentially, $\gamma$ controls the level of cluster separation in the omics space and $\delta$ represents the strength of outcome association of the clusters. We first evaluate Models 1-3 with lower level of cluster separation $\gamma=1$ and varying outcome association $\delta=2,3,5$. Model 4 evaluates $\gamma=\delta=3$.

We compare the performance of the proposed ogClust using group LASSO + $\ell_2$ penalty with three other competing clustering methods: 1) SKM: sparse $K$-means clustering \citep{witten2010framework}, a modified $K$-means algorithm with variable selection; 2) PMBC: penalized model based clustering \citep{pan2007penalized}, an unsupervised method based on Gaussian mixture model; 3) SC: supervised clustering \citep{bair2004semi}, a post-screening clustering method. SKM and PMBC are not outcome-guided and could be sensitive to any ``clinically irrelevant'' clusters, while SC has a variable pre-screening by outcome association. To evaluate the performance of these methods, we simulate $100$ datasets with sample size $n=600$, where there are 1000 genes and three subgroups with equal size. To implement SKM and PMBC and compare with ogClust, we assign observations to the cluster with closest center (SKM) or with the highest posterior probability (PMBC), then fit linear regression with covariate $X$ and outcome $Y$ in each resulting cluster to make outcome prediction. For SC, we apply a pre-screen step to pre-select $M$ outcome associated genes before we perform K-means clustering and fit linear regression in each resulting cluster, the value of $M$ is determined by cross-validation.

The performance of these methods is evaluated by their clustering accuracy, gene selection and outcome prediction by 10-fold cross-validation. Within each fold of training/testing split, we fit each of the methods using the training set and then predict both latent subgroup label and outcome value for testing set. We compute $RMSE=\sqrt{\sum_{i=1}^n(y_i-\hat{y}_i)^2/n}$ and $R^{2}=1-S S_{r e s i d u a l} / S S_{t o t a l}=1-\frac{\sum_{i=1}^{n}\left(y_{i}-\hat{y}_{i}\right)^{2}}{\sum_{i=1}^{n}\left(y_{i}-\overline{y}_{i}\right)^{2}}$ from the 10-fold cross validation and average the results to measure the prediction error of outcome (Table \ref{table:K ARI RMSE and R2}). We also compute the average number of false positives (FPs) and false negatives (FNs) for evaluating the accuracy of feature selection (Table \ref{table:K ARI RMSE and R2}). For clustering accuracy, we compute the adjusted Rand index (ARI) \citep{hubert1985comparing}, which has 0 expectation when clustering is random and bounded by 1 with perfect partition, to measure the consistency of predicted subgroup label with true latent subgroup index (Table \ref{table:K ARI RMSE and R2}). 

Table \ref{table:K ARI RMSE and R2} shows results of ogClust compared to SKM, PMBC and SC under the four simulation settings. To measure clustering performance, PMBC and SKM identifies $K=3$ clusters in 100 and 37-38 of the 100 simulations but since the algorithm has no outcome guidance, they mostly obtain clinically irrelevant clusters and have ARI= 0.04-0.16 when compared with the three true outcome-associated clusters. SC pre-selects outcome-associated gene features to perform clustering and generates improved ARI=0.35-0.41, but the method identifies $K=2$ clusters for all simulations. In contrast, ogClust identifies $K=3$ clusters for 98-99 out of 100 simulations and produces ARI=0.86-0.91 for Model 2-4. For the weak signal Model 1, ogClust identifies $K=3$ clusters for 37 of the 100 simulations and the ARI reduces to 0.45. When evaluating gene selection, PMBC misses majority of the first 15 true clustering genes (8.7-11.1 FNs) and both SKM and PMBC add many false positives (776.1-813.8 FPs for SKM and 87.0-100.3 FPs for PMBC). SC contains outcome association gene selection but still misses 4.8-8.7 FNs and adds 17.5-88.0 FPs. In contrast, ogClust almost does not miss true clustering genes (FN=0 for Model 2-4 and FN=3 for Model 1) and only adds $\sim$14 false positives. For outcome prediction result, ogClust generates the lowest RMSE and the highest $R^2$, showing better clinical relevance of produced disease subtypes. In summary, SKM and PMBC are vulnerable to miss clinically relevant clusters and related predictive genes. SC only modestly improves in detecting outcome-associated genes and clusters, and the two-stage approach reduces performance and rigor of inference. ogClust outperforms the three methods in clustering accuracy, gene selection and clinical outcome prediction. 

Table S1 and S2 show the simulation results when there is a stronger and weaker signal in $\boldsymbol{G}_{\mathcal{A}_2}$ compared with $\boldsymbol{G}_{\mathcal{A}_1}$ respectively. When the signal in $\boldsymbol{G}_{\mathcal{A}_2}$is stronger, SKM and PMBC are dominated by $\boldsymbol{G}_{\mathcal{A}_2}$ and returns clinically irrelevant clusters with ARI=0. When the signal in $\boldsymbol{G}_{\mathcal{A}_2}$is weaker, SKM and PMBC performs slightly better in identifying the three outcome associated clusters and outcome prediction with higher ARI and $R^2$. However, the expression of $\boldsymbol{G}_{\mathcal{A}_2}$ has little influence on the performance of SC and ogClust. Overall, ogClust performs consistently the best among all the simulation settings.

\begin{table}
\caption{Comparison of sparse $K$-means (SKM), penalized model based clustering (PMBC), supervised clustering (SC) and outcome-guided clustering (ogClust) under four simulation model settings with 600 observations and 2 baseline covariates, 1000 genes and 100 repetitions.}
\label{table:K ARI RMSE and R2}
\begin{center}
\begin{tabular}{ccccccccc}
\hline
Methods & \multicolumn{3}{c}{Estimated K} & ARI & \multicolumn{2}{c}{Selected Genes} & \multicolumn{2}{c}{Outcome} \\
\cline{2-4} \cline{6-7} \cline{8-9}
& 2 & 3 & $>3$ & & FPs & FNs & RMSE & $R^2$ \\
\hline
\multicolumn{5}{l}{Model 1: $\gamma=1;\delta=2$}\\
SKM & 100 & 0 & 0 & 0.04 & 776.1 & 1.9 & 1.93 & 0.25 \\
PMBC & 82 & 6 & 12 & 0.08 & 88.0 & 11.1 & 1.93 & 0.24\\
SC & 100 & 0 & 0 & 0.35 & 41.3 & 4.8 & 1.58 & 0.48 \\
ogClust & 62 & 37 & 1 & 0.45 & 5.9 & 3.0 & 1.55 & 0.51\\
\hline
\multicolumn{5}{l}{Model 2: $\gamma=1;\delta=3$}\\
SKM & 100 & 0 & 0 & 0.04 & 776.1 & 1.9 & 2.65 & 0.15 \\
PMBC & 82 & 11 & 7 & 0.10 & 87.0 & 10.2 & 2.67 & 0.14 \\
SC & 100 & 0 & 0 & 0.36 & 33.4 & 4.9 & 2.08 & 0.47\\
ogClust & 2 & 98 & 0 & 0.86 & 14.6 & 0.0 & 1.90 & 0.56\\
\hline
\multicolumn{5}{l}{Model 3: $\gamma=1; \delta=5$ }\\
SKM & 100 & 0 & 0 & 0.04 & 776.1 & 1.9 & 4.20 & 0.05 \\
PMBC & 74 & 11 & 15 & 0.09 & 100.3 & 10.2 & 4.22 & 0.05\\
SC & 100 & 0 & 0 & 0.36 & 37.9 & 4.9 & 3.20 & 0.46\\
ogClust & 0 & 99 & 1 & 0.91 & 14.5 & 0.0 & 2.70 & 0.61\\
\hline
\multicolumn{5}{l}{Model 4: $\gamma=3; \delta=3$ }\\
SKM & 100 & 0 & 0 & 0.05 & 813.8 & 1.6 & 2.61 & 0.15\\
PMBC & 83 & 5 & 12 & 0.16 & 96.7 & 8.7 & 2.64 & 0.15\\
SC & 100 & 0 & 0 & 0.41 & 17.5 & 5.0 & 2.01 & 0.48\\
ogClust & 1 & 99 & 0 & 0.88 & 12.0 & 0.0 & 1.75 & 0.63\\
\hline
\end{tabular}
\end{center}
\end{table}

\subsection{Robust Estimation under Outliers or Heavy-tailed Errors}
\label{s:robust_sim}
To compare the performance of robust methods in guarding against outliers or violation of Gaussian mixture assumption, we perform simulation in the following settings:
\begin{itemize}
    \item Setting A: The error term in the outcome association model is randomly drawn from standard normal distribution; normal assumption is not violated.
    \item Setting B: 10\% of the observations are outliers and the error term is randomly drawn from unif(min-10,max+10).
    \item Setting C: The error term is randomly drawn from heavy-tailed lognormal distribution with log-mean 0 and log-standard deviation 1.
\end{itemize}
 The simulation scheme follows Model 2 in Section \ref{s:sim}, except that in step 5, the generation of outcome $Y$ varies according to the different settings above. Under each setting, we compare the performance of ogClust, ogClust-Huber, ogClust-adHuber, and ogClust-median-truncation. Models are fit in the training data and tested in the testing data where four measures, i.e. RMSE, $R^2$, ARI and FNs, are calculated. We tune the number of selected genes by altering the parameter $\lambda$. The analysis above is performed on 100 sets of training and testing data such that we can obtained smooth curves capturing the trend of the four measures against the varying number of selected genes. 

\begin{figure}[!b]
	\noindent\makebox[\textwidth]{\includegraphics[scale=0.62]{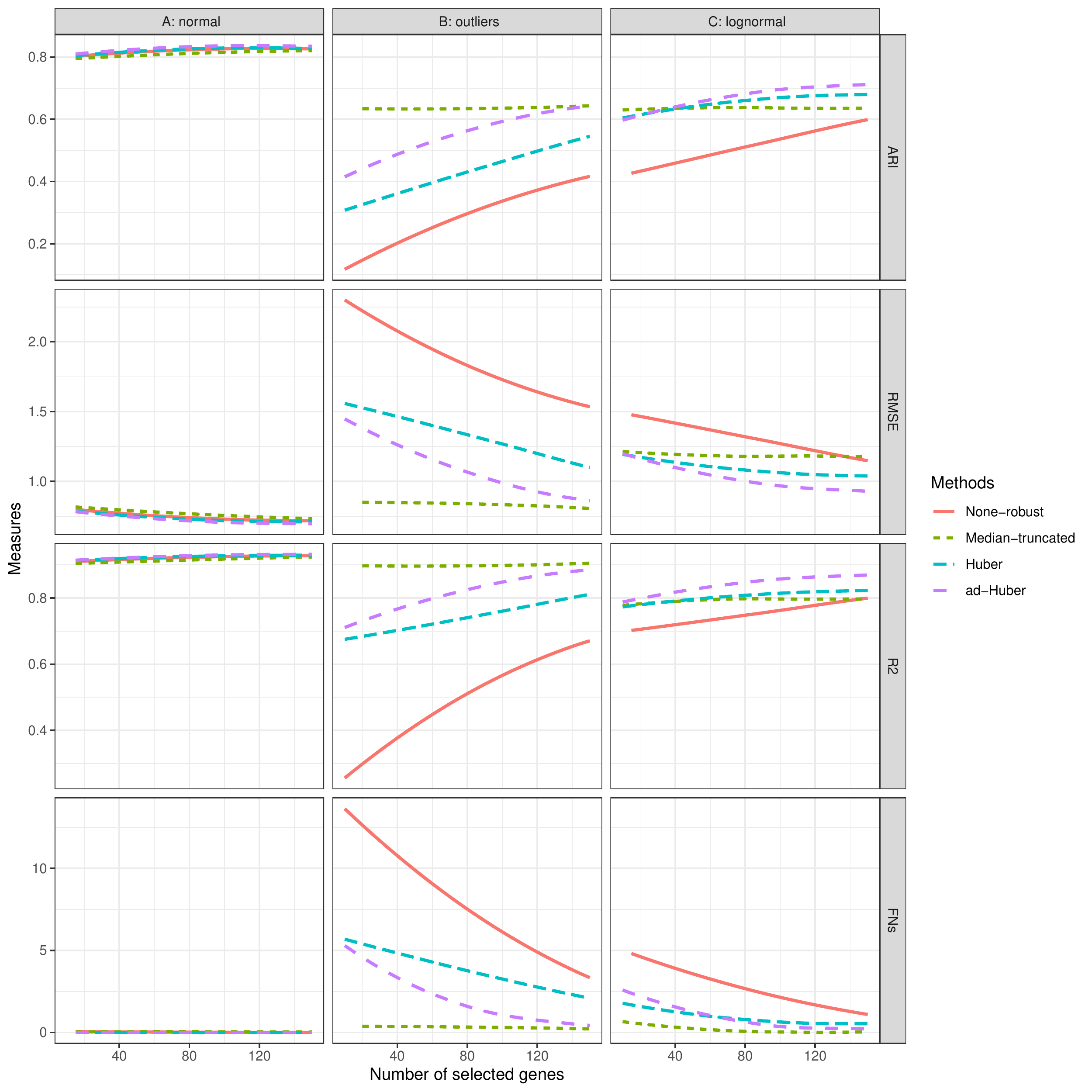}}
	\caption{Comparison of ogClust and three robust ogClust methods under settings A: error term is randomly drawn from standard normal distribution, setting B: 10\% of the observations are outliers, and setting C: error term is randomly drawn from heavy-tailed lognormal distribution. We compare RMSE, $R^2$, ARI and FNs (y-axis) vs number of genes selected in each setting (x-axis).}
    \label{fig:robust}
\end{figure}

As shown in Figure \ref{fig:robust}, in the first column when the normal assumption is satisfied, the non-robust ogClust model performs the best and the three robust methods have only very slightly worse performance. This shows that robust estimation methods only minimally reduce efficiency when Gaussian mixture assumption is true. On the other hand, when Gaussian assumption is violated in the second and the third columns, the three robust methods greatly outperform the original model. ogClust-adHuber consistently outperforms ogClust-Huber with fixed cutoff. Compared to median truncation, ogClust-adHuber performs better for heavy-tailed error term but slightly worse with existence of outliers. 
ogClust-median-truncation can quickly capture the outcome associated DE genes with relatively low number of selected genes, but it performs worse than ogClust-adHuber in setting C because of the bias in parameter estimates. Since ogClust-adHuber outperforms ogClust-huber overall and performs well in most settings, it is recommended for general applications and will be evaluated in real data in Section 4.

\subsection{Simulation to Evaluate ogClust for Survival Outcome}
\label{s:survival_sim}
The simulation scheme is the same as in Section \ref{s:sim}, except that in step 5, survival outcome is generated as follows: given subgroup index $Z$, survival time $Y$ follows AFT model with log-logistic distribution, i.e. $(log(Y)|Z=k) = \beta_{0 k} + X\beta + W\sigma$, where W $\sim$ standard logistic distribution and $\sigma=0.5$. We set the end of follow-up time to be 100, any time that is greater than 100 is right-censored. 

We evaluate the performance under four settings: (A) $\gamma=1$ and $\delta=1$, (B) $\gamma=3$ and $\delta=1$, (C) $\gamma=1$ and $\delta=2$, and (D) $\gamma=3$ and $\delta=2$, representing varying level of cluster separation (reflected by $\gamma$) and outcome association ($\delta$). Similar to Section \ref{s:robust_sim}, we compare the performance of SKM, PMBC, SC and ogClust in terms of RMSE, $R^2$, ARI and FNs under each setting. Models are evaluated in 100 sets of simulated training and testing data. We vary the number of selected genes by tuning the penalty parameter $\lambda$ and obtain smooth curves representing the trend of the four measures against the varying number of selected genes.

As the result shown in Figure S2, SKM and PMBC have the lowest ARI, highest RMSE, lowest $R^2$ and highest FNs among all four settings because they lack outcome guidance. SC has improved the four measures when compared with SKM and PMBC, and ogClust consistently outperforms the other three methods in all simulation settings.

\section{Real Application}
\label{s:real}
We apply the ogClust model to a lung disease transcriptomic dataset with $n=319$ patients.  Gene expression data are collected from Gene Expression Omnibus (GEO) GSE47460 and clinical information obtained from Lung Genomics Research Consortium (\url{https://ltrcpublic.com/}). Majority of patients were diagnosed by two most representative lung disease subtypes: chronic obstructive pulmonary disease (COPD) and interstitial lung disease (ILD). COPD is a progressive lung disease caused by the repeated exposure to a noxious agent and is classified by symptoms, airflow obstruction and exacerbation history. ILD is a loosely defined group of patients characterized by changes in the interstitium of the lung, causing pulmonary restriction and impaired gas exchange. Current clinical classification criteria of the subtypes evolve over time and are debatable. They often fail to accommodate patients with atypical features, who are left unclassified. The current criteria also fail to reflect advances of high-throughput mRNA expression techniques to improve understanding and interpretation of the disease subtypes. In this section, we utilize the standardized form of a patient's forced expiratory score (FEV1\%prd), a person's measured FEV1 normalized by the predicted FEV1 with healthy lung, as the clinical outcome $Y$ to guide the disease subtyping. Age, gender and BMI are included as covariates $X$ in the ogClust model.

\begin{figure}[!b]
	\noindent\makebox[\textwidth]{\includegraphics[scale=0.75]{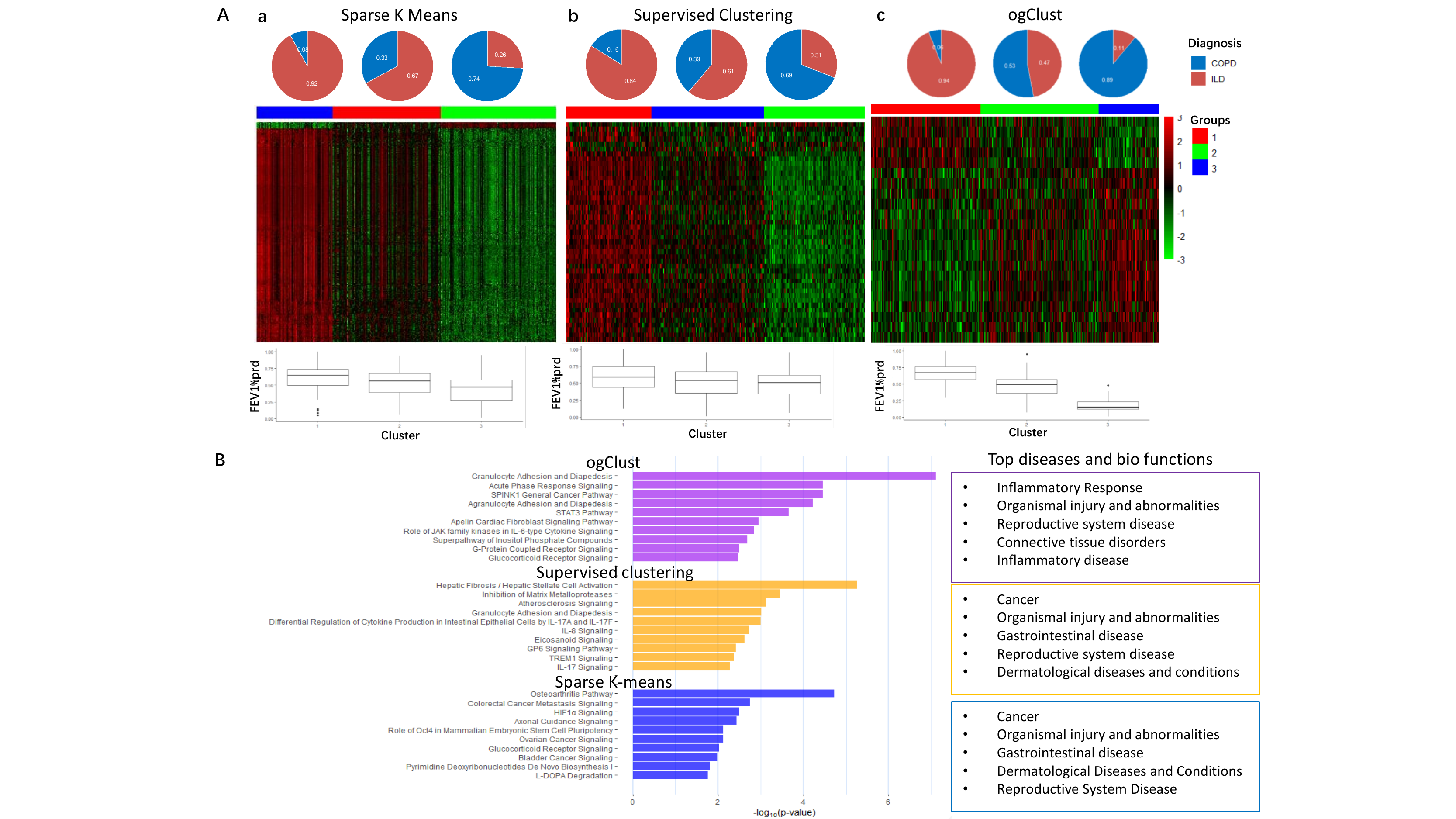}}
	\caption{(A) Pie chart of clinical diagnosis (top), heatmap of expression of selected genes (middle), and boxplot of outcome FEV1\%prd (bottom) in each cluster for (a) SKM ,(b) SC, and (c) ogClust. (B) Enriched pathways and top disease annotations of the selected genes for SKM, SC and ogClust.}
    \label{fig:real}
\end{figure}

Similar to simulations, we apply ogClust and compare with two existing methods, sparse $K$-means and supervised clustering. Data are first preprocessed by conventional procedures following an earlier publication \citep{kim2015integrative}. Non-expressed genes (mean expression in the lower 50 percentile) are filtered and top informative genes (genes with the largest variance) are selected for analysis. Table \ref{table:real data result} shows the result when setting the number of subgroups $K=3$ (see Figure S3 for analysis of justifying selection of $K$), and using the top 500, 1000 and 2000 pre-filtered genes (by the largest variance) in the comparison. Since, unlike in simulations, the underlying true class labels $Z$ are unknown, we benchmark the clustering performance in several measures. We compare the outcome prediction error using RMSE and ${R^2}$ and evaluate $p$-value of the association between subgroups and the FEV1\%prd outcome by Kruskal-Wallis test. We also show the number of selected genes, which has at least one non-zero $\hat{\gamma}_{j k}$ ($1\leq k\leq K$), used to characterize the disease subtypes. The result in Table \ref{table:real data result} shows that ogClust identifies disease subtypes with better association with clinical outcome with smaller number of genes compared to sparse $K$-means and supervised clustering. For example, when the top 2000 pre-filtered genes are used, ogClust selects 22 genes to define three disease subtypes that explain FEV1\%prd outcome with $R^2=0.350$ and association $p=1.84 \times 10^{-57}$. In contrast, sparse $K$-means needs 253 genes to reach $R^2=0.055$ and $p=5.11\times 10^{-7}$. Although supervised clustering also aims to detect subtypes associated with outcome, it only improves slightly from sparse $K$-means with $R^2$=0.058 and $p=2.11 \times 10^{-8}$. Compared with ogClust, ogClust-adHuber better explains outcome with $R^2=0.455$, and has relatively lower association with $p=9.49 \times 10^{-24}$. Figure \ref{fig:real}A shows the clinical diagnosis (piechart above), expression of the selected genes (heatmap in the middle), distribution of outcome (boxplot below) for each method. For the three clusters identified by ogClust, one cluster is almost purely COPD (blue bar), one cluster is almost purely ILD (red bar) and one cluster in between with mixed COPD and ILD (green bar). The result indicates existence of a COPD/ILD intermediate subtype of patients that have distinct molecular expression pattern and FEV1\%prd clinical outcome. SKM and SC, however, identify three clusters with more mixed diagnosis of COPD and ILD and are dominated by non-outcome-related genes. We next evaluate the enriched pathways and canonical functions using Ingenuity Pathway Analysis (IPA) tool. To account for the randomness of gene selection, we repeat the analysis in 500 bootstrapped datasets and select the top 200 most frequently selected genes as our final input gene list for IPA. As shown in Figure \ref{fig:real}B, the genes selected by ogClust are more significantly enriched in pathways associated with immune responses and organismal injury, while other methods select genes largely irrelevant to lung disease (e.g. cancer and dermatological diseases).

\begin{table}
\caption{Comparison of sparse K-means (SKM), supervised clustering (SC), outcome-guided clustering (ogClust), and ogClust with adaptive-Huber loss (ogClust-adHuber) when applied to the lung disease transcriptomic dataset. We set the number of subgroups $K$ equals 3, top 500, 1000, and 2000 genes are used. RMSE and $R^2$ measure outcome prediction performance. Kruskal-Wallis test measures whether outcome is associated with the clusters. Fisher's exact test measures whether subgroup label is consistent with the clinical diagnosis.} 
\label{table:real data result}
\begin{center}
\resizebox{1\textwidth}{!}{
\begin{tabular}{cccccccc}
\hline
K & Total number& Methods & RMSE & $R^2$ & Kruskal-Wallis & Genes & Fisher's exact \\
  & of genes & & & & test & selected & test\\
\hline
 &   & SKM & 0.208 & 0.060 & $7.28\times 10^{-5}$ & 218 & $1.34 \times 10^{-9}$\\
3 & 500 & SC & 0.203 & 0.101 & $1.36\times 10^{-7}$ & 70 & $3.65 \times 10^{-24}$\\
 &   & ogClust & 0.189 & 0.226 & $7.21 \times 10^{-56}$ & 33 & $ 2.81 \times 10^{-41}$ \\
 &   & ogClust-adHuber & 0.168 & 0.386 & $2.24\times 10^{-47}$ & 11 & $1.12 \times 10^{-18}$ \\
\hline
  &   & SKM & 0.209 & 0.052 & $1.79 \times 10^{-6}$ & 172 & $1.27 \times 10^{-7}$ \\
3 &1000 & SC & 0.204 & 0.086 & $2.31\times 10^{-5}$ & 60 & $4.43 \times 10^{-21}$ \\
  &   & ogClust & 0.186 & 0.249 & $7.62\times 10^{-56}$ & 40 & $ 8.05 \times 10^{-41}$\\
  &   & ogClust-adHuber & 0.161 & 0.432 & $1.00 \times 10^{-57}$ & 25 & $1.87 \times 10^{-31}$ \\
\hline
 &   & SKM & 0.208 & 0.055 & $5.11\times 10^{-7}$ & 253 &  $ 4.16 \times 10^{-23}$\\
3 &2000 & SC & 0.207 & 0.058 & $2.11\times 10^{-8}$ & 45 & $ 8.52 \times 10^{-14}$\\
 &   & ogClust & 0.173 & 0.350 & $1.84 \times 10^{-57}$ & 22 & $ 8.56 \times 10^{-34}$\\
 &   & ogClust-adHuber & 0.158 & 0.455 & $9.49\times 10^{-24}$ & 24 & $5.51 \times 10^{-43}$ \\
\hline
\end{tabular}
}
\end{center}
\end{table}

\section{Conclusion}
\label{s:conclusion}
In this paper, we propose a unified outcome-guided clustering (ogClust) framework for disease subtyping from omics data. ogClust links disease subtyping model and outcome association model through a latent cluster label $Z$. From extensive simulations and a real data application on lung disease transcriptomic data, we demonstrate the ability of ogClust to identify outcome associated clusters (disease subtypes)  that are otherwise easily masked by other facets of clinically irrelevant cluster structure. Additionally, ogClust is immediately applicable to future patients to predict their disease subtypes. Unlike hard (deterministic) assignment in hierarchical clustering or $K$-means, the prediction is a soft assignment with classification probability, reflecting confidence of subtyping prediction of each patient.

As mentioned in the Introduction section, the concept of outcome-guided clustering is novel in the field. It involves both supervised and unsupervised components in the framework but differs from classical clustering or classification problems. It should not be confused with two types of semi-supervised machine learning, where mixing of labeled and unlabeled data are trained or constrained prior knowledge is imposed in clustering. To some extent, it is similar to latent class models in outcome association, but the latter model cannot provide latent class assignment for future observations, while ogClust model can predict disease subtypes for precision medicine purpose.

In the current ogClust model, omics data $\boldsymbol{G}$ from a single source are used to characterize the subtype $Z$ and covariates $\boldsymbol{X}$ do not contribute to clustering. Integration of multi-source of data (e.g. multiple transcriptomic studies or a single study with multi-omics data) requires more careful modeling for each problem setting and will be a future direction.

ogClust parameter estimation is implemented via a modified EM algorithm and thus provides fast computing for high-dimensional data. In the lung disease example, the model fitting can be finished in 2.17 minutes using 1 core (Intel Xeon 6130) for $n=319$ patients, $q=2000$ genes and $p=3$ covariates. To select tuning parameters $K$ and $\lambda$ by BIC, multiple runs are necessary. An R package is freely available on \url{https://github.com/liupeng2117/ogClust}, along with all data and code to reproduce results in this paper.  


\bibliographystyle{apalike} 
\bibliography{references}



\label{lastpage}

\end{document}


\title{Supplementary Materials for ``Outcome-Guided Disease Subtyping for High-Dimensional Omics Data"\\[10pt]

{\large Peng Liu, Yusi Fang, Zhao Ren, Tang Lu and George C. Tseng}}

\maketitle


\beginsupplement

\begin{figure}[!b]
	\noindent\makebox[\textwidth]{\includegraphics[scale=0.6]{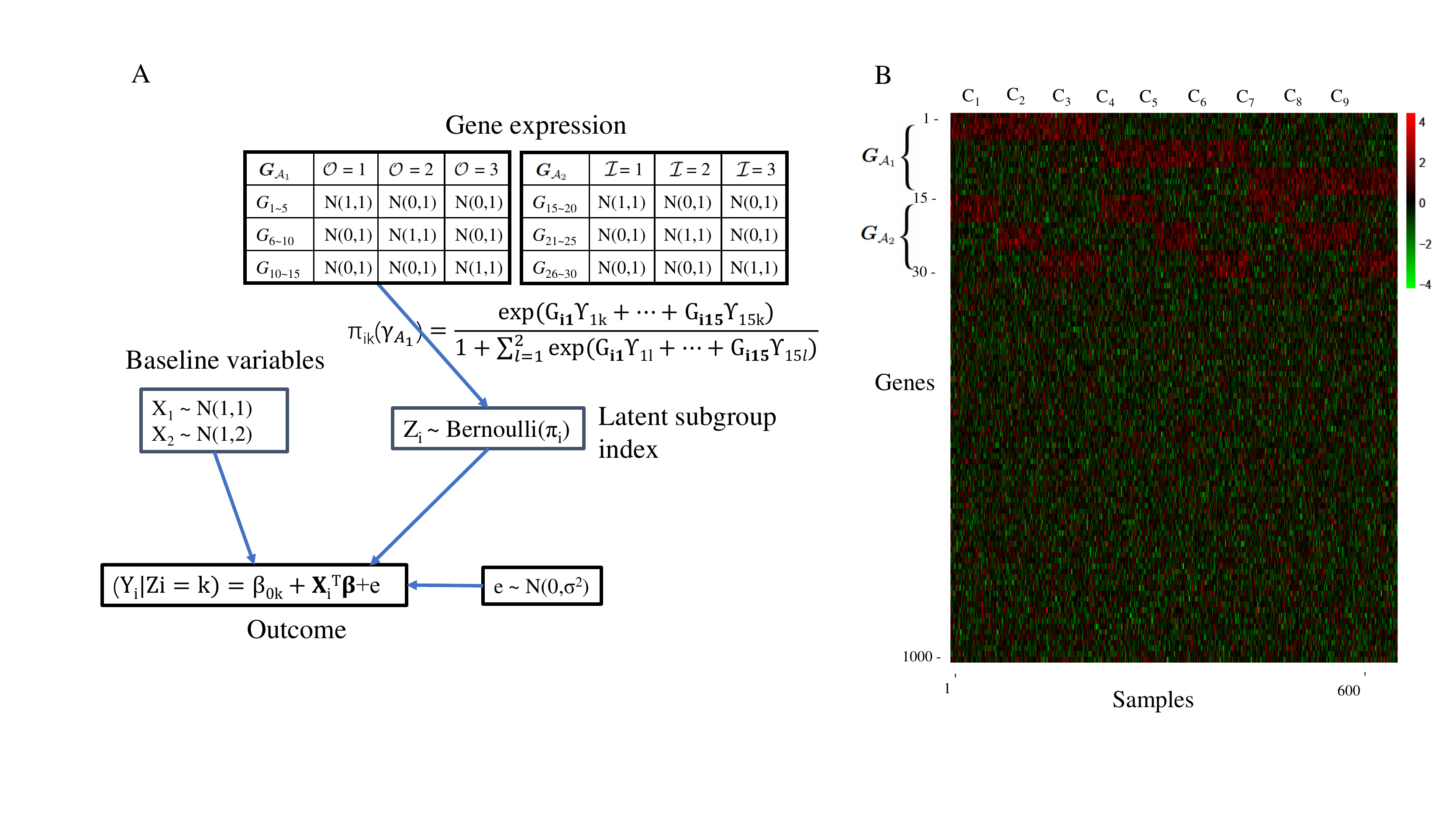}}
	\caption{(A) Data generation scheme. $\mathcal{O}=\{1,2,3\}$ denotes three clusters defined by genes set $G_{\mathcal{A}_1}$, $\mathcal{A}_1 = \{ 1, \dots ,15 \}$, and $\mathcal{I}=\{1,2,3\}$ denotes another three independent clusters defined by $G_{\mathcal{A}_2}$, $\mathcal{A}_2 = \{ 16, \dots ,30 \}$. Expression of genes in $G_{\mathcal{A}_1}$ and $G_{\mathcal{A}_2}$ are generated from the distributions listed on the above table. For subject i, only $G_{\mathcal{A}_1}$ have real signals effecting $Z_i$, which is drawn from a Multinomial distribution with probability $\boldsymbol{\pi}_i=\{\pi_{i1},\pi_{i2},1-\pi_{i1}-\pi_{i2}\}$. Baseline variables $X_1$ and $X_2$ are generated from $N(1,1)$ and $N(1,2)$ respectively. Given $X_i$, $G_i$ and $Z_i$, the outcome $Y_i$ is generated finally. (B) Heatmap of the expression of 1000 genes across samples. A total of nine subgroups $C_1$, ..., $C_9$ are jointly defined by genes sets $G_{\mathcal{A}_1}$ and $G_{\mathcal{A}_2}$.} 
    \label{fig:simulation}
\end{figure}

\begin{figure}[!b]
	\noindent\makebox[\textwidth]{\includegraphics[scale=0.7]{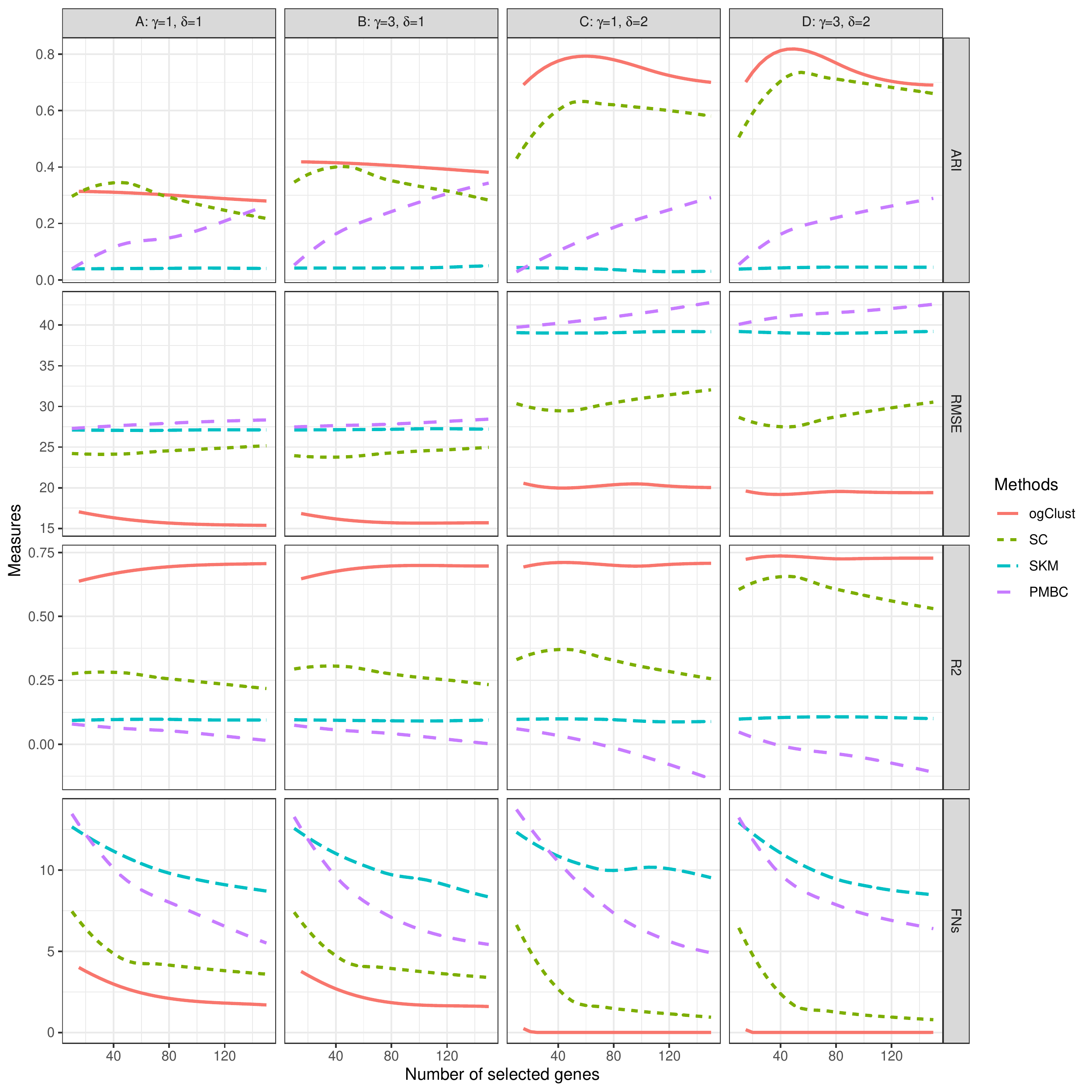}}
	\caption{Comparison of ogClust and SC,SKM and PMBC under four simulation settings with survival outcome. We compare RMSE, $R^2$, ARI and FNs (y-axis) vs number of genes selected in each setting (x-axis).} 
    \label{fig:simulation}
\end{figure}

\begin{figure}[!b]
	\noindent\makebox[\textwidth]{\includegraphics[scale=0.8]{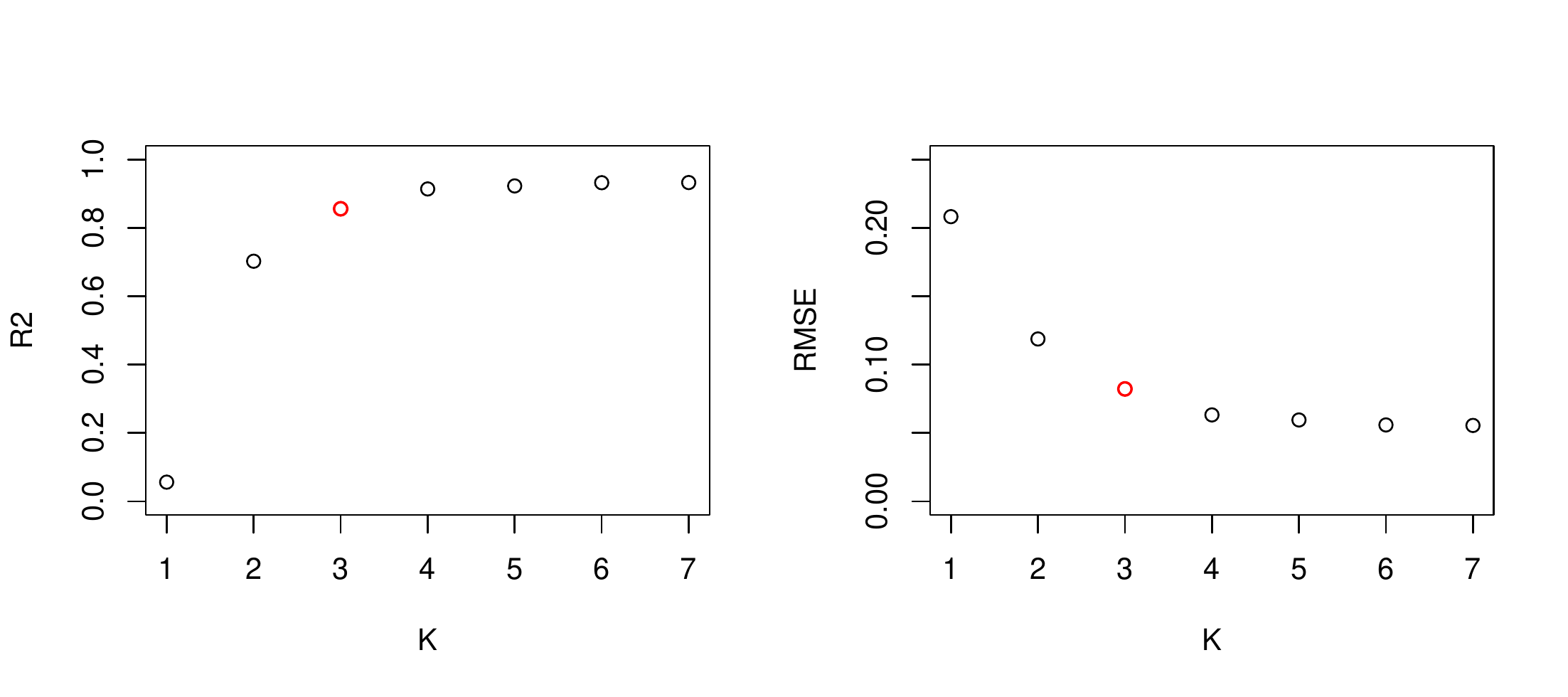}}
	\caption{Plot of (A) $R^2$ and (B) RMSE against the number of clusters.} 
    \label{fig:simulation}
\end{figure}

\begin{table}
\caption{Comparison of sparse k-means (SKM), penalized model based clustering (PMBC), supervised clustering (SC) and outcome-guided clustering (ogClust) under four simulation model settings with 600 observations and 2 baseline covariates, 1000 genes and $G_{j|j \in \mathcal{A}_2} \sim N(3, 1)$ in 100 repetitions.} 
\label{table:K ARI RMSE and R2}
\begin{center}
\begin{tabular}{ccccccccc}
\hline
Methods & \multicolumn{3}{c}{Estimated K} & ARI & \multicolumn{2}{c}{Selected Genes} & \multicolumn{2}{c}{Outcome} \\
\cline{2-4} \cline{6-7} \cline{8-9}
& 2 & 3 & $>3$ & & FPs & FNs & RMSE & $R^2$ \\
\hline
\multicolumn{5}{l}{Model 1: $\gamma=1;\delta=2$}\\
SKM & 37 & 63 & 0 & 0.00 & 286.2 & 9.9 & 1.94 & 0.24 \\
PMBC & 0 & 97 & 3 & 0.00 & 78.7 & 13.4 & 1.93 & 0.24\\
SC & 100 & 0 & 0 & 0.35 & 45.3 & 4.8 & 1.59 & 0.46 \\
ogClust & 41 & 59 & 0 & 0.45 & 5.8 & 3.0 & 1.55 & 0.51\\
\hline
\multicolumn{5}{l}{Model 2: $\gamma=1;\delta=3$}\\
SKM & 37 & 63 & 0 & 0.00 & 286.2 & 9.9 & 2.68 & 0.14 \\
PMBC & 0 & 95 & 5 & 0.00 & 85.3 & 13.4 & 2.68 & 0.13\\
SC & 100 & 0 & 0 & 0.36 & 28.5 & 5 & 2.13 & 0.45\\
ogClust & 2 & 98 & 0 & 0.86 & 14.6 & 0 & 1.90 & 0.55\\
\hline
\multicolumn{5}{l}{Model 3: $\gamma=1; \delta=5$ }\\
SKM & 37 & 63 & 0 & 0.00 & 286.2 & 9.9 & 4.25 & 0.05 \\
PMBC & 0 & 98 & 2 & 0.00 & 92.0 & 13.2 & 4.27 & 0.04\\
SC & 100 & 0 & 0 & 0.36 & 32.4 & 4.9 & 3.25 & 0.42\\
ogClust & 1 & 99 & 0 & 0.91 & 14.4 & 0 & 2.72 & 0.61\\
\hline
\multicolumn{5}{l}{Model 4: $\gamma=3; \delta=3$ }\\
SKM & 38 & 62 & 0 & 0.00 & 406.1 & 8.5 & 2.67 & 0.14\\
PMBC & 0 & 100 & 0 & 0.00 & 82.0 & 13.8 & 2.67 & 0.13\\
SC & 100 & 0 & 0 & 0.41 & 17.5 & 5 & 2.20 & 0.41\\
ogClust & 2 & 98 & 0 & 0.88 & 12.1 & 0 & 1.74 & 0.63\\
\hline
\end{tabular}
\end{center}
\end{table}

\begin{table}
\caption{Comparison of sparse k-means (SKM), penalized model based clustering (PMBC), supervised clustering (SC) and outcome-guided clustering (ogClust) under four simulation model settings with 600 observations and 2 baseline covariates, 1000 genes and $G_{j|j \in \mathcal{A}_2} \sim N(0.5, 1)$ in 100 repetitions.} 
\label{table:K ARI RMSE and R2}
\begin{center}
\begin{tabular}{ccccccccc}
\hline
Methods & \multicolumn{3}{c}{Estimated K} & ARI & \multicolumn{2}{c}{Selected Genes} & \multicolumn{2}{c}{Outcome} \\
\cline{2-4} \cline{6-7} \cline{8-9}
& 2 & 3 & $>3$ & & FPs & FNs & RMSE & $R^2$ \\
\hline
\multicolumn{5}{l}{Model 1: $\gamma=1;\delta=2$}\\
SKM & 100 & 0 & 0 & 0.05 & 794.0 & 1.6 & 1.94 & 0.24 \\
PMBC & 73 & 1 & 26 & 0.33 & 182.7 & 4.4 & 1.93 & 0.24\\
SC & 100 & 0 & 0 & 0.35 & 47.9 & 4.8 & 1.59 & 0.48 \\
ogClust & 66 & 31 & 3 & 0.45 & 14.0 & 3.3 & 1.55 & 0.51\\
\hline
\multicolumn{5}{l}{Model 2: $\gamma=1;\delta=3$}\\
SKM & 100 & 0 & 0 & 0.05 & 794.0 & 1.6 & 2.66 & 0.13 \\
PMBC & 74 & 0 & 26 & 0.30 & 172.0 & 5.9 & 2.68 & 0.13\\
SC & 100 & 0 & 0 & 0.36 & 51.0 & 4.8 & 2.09 & 0.47\\
ogClust & 2 & 97 & 1 & 0.86 & 21.3 & 0.1 & 1.90 & 0.56\\
\hline
\multicolumn{5}{l}{Model 3: $\gamma=1; \delta=5$ }\\
SKM & 100 & 0 & 0 & 0.05 & 794.0 & 1.6 & 4.22 & 0.05 \\
PMBC & 69 & 1 & 30 & 0.28 & 171.8 & 6.5 & 4.24 & 0.04\\
SC & 100 & 0 & 0 & 0.36 & 47.5 & 4.8 & 3.21 & 0.46\\
ogClust & 0 & 100 & 0 & 0.91 & 5.1 & 0.1 & 2.70 & 0.61\\
\hline
\multicolumn{5}{l}{Model 4: $\gamma=3; \delta=3$ }\\
SKM & 100 & 0 & 0 & 0.06 & 769.0 & 2.0 & 2.62 & 0.15\\
PMBC & 77 & 1 & 22 & 0.33 & 187.8 & 6.4 & 2.64 & 0.15\\
SC & 100 & 0 & 0 & 0.41 & 17.3 & 5.0 & 2.02 & 0.51\\
ogClust & 0 & 100 & 0 & 0.88 & 2.5 & 0.1 & 1.75 & 0.63\\
\hline
\end{tabular}
\end{center}
\end{table}


\label{lastpage}

